\documentclass[journal,10pt,a4paper]{IEEEtran}
\pdfoutput=1
\IEEEoverridecommandlockouts
\usepackage{amsbsy}
\usepackage{amsthm}
\usepackage{multirow}
\usepackage{epsfig}
\usepackage{subfigure}
\usepackage{graphics}
\usepackage{multicol}
\usepackage{amsmath,amssymb,amsfonts}
\usepackage{epstopdf}
\usepackage{balance}
\usepackage{xcolor}
\usepackage{cite}
\usepackage{color}
\usepackage{mathrsfs}
\usepackage{textcomp}
\usepackage{subfloat}
\usepackage{booktabs}
\usepackage{bm}
\usepackage{filecontents}
\usepackage{acronym}
\usepackage{upgreek}
\usepackage {ragged2e}
\usepackage{diagbox}
\usepackage{hyperref}
\usepackage[ruled, boxed, linesnumbered]{algorithm2e}

\graphicspath{ {figures/}}

\acrodef{cas}[CAS]{communication-assisted sensing}
\acrodef{sac}[SAC]{sensing-assisted communication}
\acrodef{snc}[S\&C]{sensing and communications}
\acrodef{isac}[ISAC]{integrated sensing and communications}
\acrodef{rd}[RD]{rate-distortion}
\acrodef{iid}[i.i.d.]{independently and identically distributed}
\acrodef{mse}[MSE]{mean squared error}
\acrodef{mmse}[MMSE]{minimum mean squared error}
\acrodef{glm}[GLM]{Gaussian linear model}
\acrodef{mi}[MI]{mutual information}
\acrodef{sp}[SP]{sensing process}
\acrodef{cp}[CP]{communication process}
\acrodef{wrt}[w.r.t.]{with respect to}
\acrodef{sct}[SCT]{source-channel separation theorem}
\acrodef{qos}[QoS]{quality of service}
\acrodef{snr}[SNR]{signal to noise ratio}
\acrodef{sw}[SW]{separated S\&C waveforms}
\acrodef{dw}[DW]{dual-functional waveform}

\newcommand{\mathept}[1]{\mathbb{E} \left[ #1 \right]}

\newcommand{\tr}[1]{\text{Tr}\left( #1 \right)}

\newcommand{\Rv}[1]{\mathsf{#1}}

\newcommand{\diag}[1]{\text{diag}\left( #1 \right)}

\newcommand{\logdet}[1]{\log \det \left( #1 \right)}

\begin{document}

\title{Communication-Assisted Sensing in 6G Networks}

\author{Fuwang Dong,~\IEEEmembership{Member,~IEEE}, Fan Liu,~\IEEEmembership{Senior Member,~IEEE}, Shihang Lu, Yifeng Xiong,~\IEEEmembership{Member,~IEEE}, \\ Qixun Zhang,~\IEEEmembership{Member,~IEEE}, Zhiyong Feng,~\IEEEmembership{Senior Member,~IEEE}, and Feifei Gao,~\IEEEmembership{Fellow,~IEEE}

\thanks{(\textit{Corresponding author: Fan Liu.})}
\thanks{Part of this paper was presented at IEEE/CIC International Conference on Communications in China (ICCC) 2023 \cite{10233679}.}
\thanks{Fuwang Dong, Fan Liu, and Shihang Lu are with the School of System Design and Intelligent Manufacturing, Southern University of Science and Technology, Shenzhen 518055, China. (email: dongfw@sustech.edu.cn; liuf6@sustech.edu.cn; lush2021@mail.sustech.edu.cn).}
\thanks{Yifeng Xiong is with the School of Information and Electronic Engineering, Beijing University of Posts and Telecommunications, Beijing 100876, China. (email: yifengxiong@bupt.edu.cn).}
\thanks{Qixun Zhang and Zhiyong Feng are with Key Laboratory of Universal Wireless Communications Ministry of Education, Beijing University of Posts and Telecommunications, Beijing 100876, China. (email: zhangqixun@bupt.edu.cn; email: fengzy@bupt.edu.cn).}
\thanks{Feifei Gao is with Department of Automation, Tsinghua University, Beijing 100876, China. (email: feifeigao@ieee.org).}
}

\maketitle

\thispagestyle{empty}
\pagestyle{empty}

\begin{abstract}
Exploring the mutual benefit and reciprocity of sensing and communication (S\&C) functions is fundamental to realizing deeper integration for integrated sensing and communication (ISAC) systems. This paper investigates a novel communication-assisted sensing (CAS) system within 6G perceptive networks, where the base station actively senses the targets through device-free wireless sensing and simultaneously transmits the estimated information to end-users. In such a CAS system, we first establish an optimal waveform design framework based on the rate-distortion (RD) and source-channel separation (SCT) theorems. After analyzing the relationships between the sensing distortion, coding rate, and communication channel capacity, we propose two distinct waveform design strategies in the scenario of target impulse response estimation. In the separated S\&C waveforms scheme, we equivalently transform the original problem into a power allocation problem and develop a low-complexity one-dimensional search algorithm, shedding light on a notable \emph{power allocation tradeoff} between the S\&C waveform. In the dual-functional waveform scheme, we conceive a heuristic mutual information optimization algorithm for the general case, alongside a modified gradient projection algorithm tailored for the scenarios with independent sensing sub-channels. Additionally, we identify the presence of both \emph{subspace tradeoff} and \emph{water-filling tradeoff} in this scheme. Finally, we validate the effectiveness of the proposed algorithms through numerical simulations.                    
\end{abstract}

\IEEEpeerreviewmaketitle

\section{Introduction}\label{Introduction}
\IEEEPARstart{T}{he} next-generation wireless networks (5G-A and 6G) are envisioned to simultaneously provide high-precision sensing capabilities and ubiquitous wireless connectivity with the help of \ac{isac} technology, which has been officially approved as one of the six key usage scenarios of 6G by the international telecommunications union (ITU) \cite{ITU2023}. \ac{isac} system is expected to attain \emph{integration gain} and \emph{coordination gain} \cite{9606831}, facilitating the \ac{snc} performance. Over the past few decades, significant research efforts have been dedicated to exploring and harnessing the \emph{integration gain}, aiming to enhance \ac{snc} performance and improve resource efficiency by the shared use of both wireless resources and hardware platforms. This line of research includes the development of dual-functional \ac{isac} waveforms, beamforming, and transmission protocols (cf.\cite{9540344,9737357,9858656}, and the reference therein). However, there exists a notable gap in the literature concerning the exploration of the \emph{coordination gain} offered by the collaboration of \ac{snc} subsystems.

\subsection{Sensing-Assisted Communication Framework}
The primary goal of the \ac{sac} framework is to establish reliable communication links supported by radar sensing capabilities, thereby significantly reducing the signaling overhead induced by frequent beam training and channel estimation in high-mobility scenarios. The approach proposed by \cite{9171304} pioneers a beam tracking and prediction method using extended Kalman filtering in vehicular networks, where the roadside unit (RSU) simultaneously tracks and communicates with vehicles via dual-functional signal. This work shows that the \ac{snc} performance within the \ac{sac} framework outperforms that of conventional feedback-based beam tracking approaches \cite{9171304}. In \cite{9246715}, the \ac{sac} framework's performance is further enhanced through advanced Bayesian inference techniques, achieving near-optimal results by a tailored message passing algorithm.

Building on the above works, the authors of \cite{9947033} address extended target scenarios in massive multi-input multi-output (MIMO) systems, presenting a dynamic predictive beamforming scheme that employs both wide and pencil-sharp beams to track communication receivers. Moving beyond the limitations of the ideal linear motion model, \cite{10061429} explores the application of \ac{sac} framework on arbitrarily shaped road to improve reliability in complex environments. Furthermore, the recent advances in the \ac{sac} systems incorporate intelligent omni-surface \cite{10226306}, deep learning-based beam selection and power allocation strategies \cite{10299722}, etc. 

Previous studies have primarily focused on enhancing communication performance with the aid of sensing capabilities. However, in 6G networks, it is well recognized that sensing services are equally important to their communication counterparts, particularly for emerging environment-ware applications \cite{9945983}. To improve the sensing performance, the sharing of sensing measurements among various devices or a fusion center through wireless communication channels has garnered extensively attention, including the applications of connected autonomous vehicle\cite{9982368,9528013}, over-the-air computation \cite{10621049}, the Internet of Things (IoT) \cite{9206051}, and extended reality (XR) \cite{8869705}. A representative use case is in artificial intelligence (AI) and edge computation/learning, where large volumes of sensing data are exchanged between edge and central nodes to enable real-time, high-accurate object recognition tasks \cite{8970161}.
 
\subsection{Communication-Assisted Sensing Framework}

\begin{figure}[!t]
	\centering
	\includegraphics[width=3.5in]{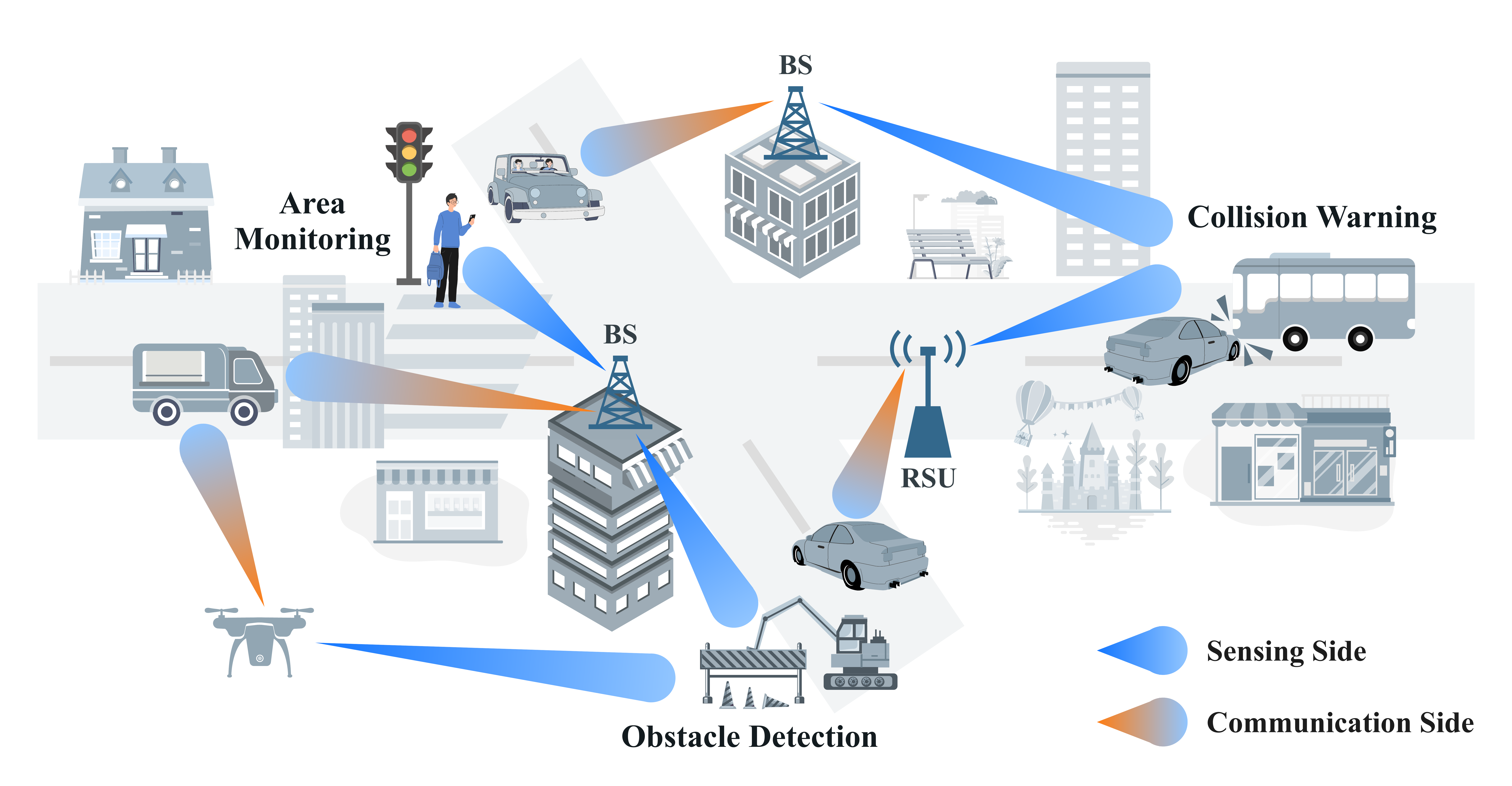}
	\caption{The applications for sensory data sharing in the proposed CAS system.}
	\label{CAS_illustration}
\end{figure}
In conventional networks, the sensing data is typically collected by numerous distributed sensors, which can be classified within the remote estimation framework \cite{gao2018optimal}. In this framework, a remote control center is responsible for data fusion, inference, and issuing control commands. However, within the 6G perceptive network \cite{9296833}, \emph{device-free} wireless sensing becomes an intrinsic capability, enabling simultaneous data acquisition and transmission on a single platform without the additional sensors. For instance \footnote{It should be highlighted that our proposed CAS framework is not limited to this specific scenario. Instead, it is applicable to a wide range of applications requiring sensory data exchange, e.g., smart factories, robotic environment monitoring, human behavior recognition, etc. Here, the meanings of to-be-estimated parameters are beyond traditional radar parameters such as range, Doppler, and azimuth.}, in the scenario depicted in Fig. \ref{CAS_illustration}, a base station (BS) or RSU with favorable visibility illuminates the targets and captures observations, constituting the \ac{sp}. The BS then transmits the sensory data acquired during this sensing stage to the end-users, completing the \ac{cp}. We refer to this combined process as \ac{cas} framework in ISAC systems, distinguishing it from conventional sensor-based networks. Additionally, we define the discrepancy between the ground truth of the targets and their reconstruction at the end-users as \emph{CAS distortion} for illustration convenience.          

The CAS distortion is evidently influenced by both the estimation distortion in the \ac{sp} and the recovery distortion in the \ac{cp}. This coupling between \ac{snc} processes manifests in two key aspects. First, \ac{snc} performance is determined by system resource management (e.g., power, bandwidth, waveform), as both processes are executed on the same hardware platform. Second, a substantial number of bits are required to represent sensing data (e.g., with uplink sensing data rates potentially reaching several gigabits per second \cite{9687468,9349624}), necessitating sufficient communication channel capacity to transmit this data reliably \footnote{From the perspective of information theory, the achievable distortion is inversely proportion to the channel capacity.}. If the BS prioritizes resource allocation for the \ac{sp} to achieve the most accurate measurements possible, it may compromise data reconstruction accuracy at the user due to limitations in communication channel capacity. This leads to a critical question addressed in this paper: \emph{How can we achieve an optimal performance tradeoff between \ac{snc} processes to minimize the CAS distortion?} 

The similar work related to the above \ac{cas} process is referred to remote source coding (also known as the \textit{chief estimation officer (CEO)} problem), where the BS can only access noisy observations rather than the original source information \cite{490552,6915877}. Two primary approaches in remote source coding are the compress-and-estimate (CE) and estimate-and-compress schemes (EC) schemes. In the CE scheme, the BS directly compresses the received observations and transmits the raw data to the user, with estimation procedures being conducted after compression. By contrast, in the EC scheme, the BS initially estimates the parameters from the observations and then transmits the compressed estimation results to the user according to the available rate. It is widely recognized that the optimal tradeoff between the information rate and the expected distortion is achieved through the EC strategy \cite{9387338}. {In light of this, we also adopt an EC scheme, also known as task-based quantization \cite{8805173}, where the BS transmits estimates instead of raw data.}  
  
In ISAC systems, the waveform design problem has attracted considerable research attention. To fulfill specific requirements, various performance metrics and design criteria have been employed to optimize waveforms. These include, but not limited to, minimizing the multi-user interference \cite{8386661}, maximizing the signal to interference and noise ratio (SINR) \cite{9399825}, minimizing the Cram\'er-Rao bound for estimation task \cite{8561147}, and maximizing the \ac{mi}, where MI between the observations and the sensing channel state serves as the sensing metric \cite{7472362,10129042}. Despite these advances, most state-of-the-art waveform designs have primarily focused on mitigating mutual interference between \ac{snc} subsystems to explore integration gain. The waveform design for CAS systems, where the waveforms are strongly coupled within the EC process, has been largely overlooked. To fill this research gap, our study is dedicated to resolving the following problem: \textit{How can we strategically design a waveform to minimize CAS distortion, while considering the \ac{snc} coupling and resource budget constraints?} 
    
\subsection{Our Contributions}

In this paper, our specific attention is directed toward target response matrix (TRM) estimation, with the to-be-estimated parameters being the sensing channel itself. As a result, we may model the EC procedure as a quadratic Gaussian problem. The main contributions are summarized as follows.

\begin{itemize}
	\item First, we establish a \ac{cas} framework in 6G perceptive networks based on the theories of the \ac{rd} and the \ac{sct} in lossy data transmission. This framework clearly characterizes the relationship between the distortion, coding rate, and channel capacity. Additionally, we develop the waveform design scheme that minimizes the CAS distortion while adhering to the \ac{sct} and power budget constraints.  
   		
	\item Second, we derive the closed-form expressions of \ac{mmse} and \ac{rd} function for the scenario of TRM estimation, and reformulate the original problem as a non-convex optimization task with reverse water-filling constraints. We then propose two distinct waveform design schemes that are widely considered in \ac{isac} systems, namely, the \ac{sw} and \ac{dw} designs. 
		  
	\item Third, for the \ac{sw} design, we simplify the original matrix-valued optimization problem into a single scalar power allocation problem. Subsequently, a one-dimensional search algorithm is proposed to obtain the optimal solution, which implies a \emph{power allocation tradeoff} between \ac{snc} waveform. 
 	
	\item Forth, for the \ac{dw} design, we present a heuristic \ac{mi} maximization algorithm for the general case, as well as a modified gradient projection algorithm for the case of independent sensing sub-channels. By doing so, we highlight the presence of both \emph{subspace tradeoff} and \emph{water-filling tradeoff} in this scenario.     
	
	\item Finally, we provide numerical simulations to validate the effectiveness of the proposed algorithms and analyze the performance tradeoffs. 
\end{itemize}

The reminder of this paper is organized as follows. In Section \ref{prelimi}, we introduce the concepts of lossy data transmission including the \ac{rd} and \ac{sct} theories, and the well-studied optimal structures for \ac{mi}-optimal communication waveform and the \ac{mmse}-optimal sensing waveform. In Section \ref{SystemModel}, we establish a comprehensive \ac{cas} framework and formulate the optimization problem for the waveform design within such systems. We then present the corresponding waveform design algorithms for two typical signaling strategies, the \ac{sw} in Section \ref{sncwd} and the \ac{dw} in Section \ref{dualfuncwd}, respectively. In Section \ref{discussion}, we briefly discuss the design insights revealed from the development of the \ac{cas} system. The simulation results and the tradeoff analysis for the S\&C performance are provided in Section \ref{simulation}. Finally, we conclude this paper in Section \ref{Conclusion}.       

Notations: The uppercase normal letter $\Rv{A}$, lowercase italic letter $a$, and fraktur letter $\mathcal{A}$ denote a random variable, its realization and a set, respectively. Uppercase and lowercase bold letters $\mathbf{A}$ and $\mathbf{a}$ respectively denote a matrix and a column vector. $\textbf{I}_N$ is the $N$-dimensional identity matrix and $\textbf{1}_N$ is the $N \times 1$ vector having all-one entries. $(\cdot)^T$, $(\cdot)^*$, and $(\cdot)^H$ represent the transpose, conjugate, and complex conjugate transpose operations, respectively. $\mathept{\cdot}$ is the statistical expectation, and $\tr{\cdot}$ is the trace of a matrix. $(\cdot)^+$ is defined by $\max\{\cdot, 0\}$; $\diag{\mathbf{a}}$ stands for a diagonal matrix using the elements of $\mathbf{a}$ as its diagonal elements. 


\section{Preliminaries}\label{prelimi}
Lossy data transmission theory plays a vital role in characterizing the relationship between the distortion, coding rate, and channel capacity. Before elaborating on the \ac{cas} system design, we briefly introduce the contents of the \ac{rd} and \ac{sct} theorems, as well as the structures of \ac{mmse}-optimal sensing waveform and \ac{mi}-optimal communication waveform.

\subsection{Lossy Data Transmission} \label{preloss}
\subsubsection{Rate-Distortion Theory} In source channel coding, \emph{distortion} is defined to measure the distance between the original source and its representation, evaluating the ``goodness'' of a representation. Let us denote $\mathsf{A}^N$ as an $N$-length sequence $\mathsf{A}_1, \mathsf{A}_2, \cdots, \mathsf{A}_N$ \ac{iid} with probability density function $p_\mathsf{A}(a)$, $a \in \mathcal{A}$. For a given coding rate $R$, a $(2^{NR},N)$ code that quantizes $\mathsf{A}^N$ consists of           
\begin{itemize}
\item A encoder $f_N:\mathcal{A}^N \to \{1,2,\cdots, 2^{NR}\}$, mapping the source alphabet to an index set.
\item A decoder $g_N:\{1,2,\cdots, 2^{NR}\} \to \hat{\mathcal{A}}^N$, reproducing the estimate $\hat{\mathsf{A}}^N$ from the index set.      
\end{itemize}     
Thus, the distortion associated with the $(2^{NR},N)$ code is defined by
\begin{equation}
D = \mathept{d(\mathsf{A}^N,g_N(f_N(\mathsf{A}^N)))},
\end{equation}   
where $d(\cdot,\cdot):\mathcal{A}^N \times \hat{\mathcal{A}}^N \to \mathbb{R}^+$ is a distance metric (e.g., Hamming distortion, squared-error distortion) defined in terms of the specific requirements.  

The \ac{rd} theory characterizes the minimum information rate required to achieve a preset distortion, or equivalently, the minimum distortion achievable at a particular rate. For a given source distribution $p_\mathsf{A}(a)$ and distance metric $d$, the rate-distortion function $R(D)$ is defined by \cite{thomas2006elements}       
\begin{equation} \label{rd1}
\begin{aligned}
& R(D)=\mathop { \rm{min} } \limits_{p_{\hat{\mathsf{A}}|\mathsf{A}}:\mathbb{E}\left[d(\mathsf{A},\hat{\mathsf{A}})\right] \le D} I(\mathsf{A};\hat{\mathsf{A}}),
\end{aligned}
\end{equation}
where $I(\mathsf{A};\hat{\mathsf{A}})$ represents the \ac{mi} between the source and its representation. Note that $R(D)$ is a \emph{monotonic non-increasing convex} function, cf. \cite[Lemma 10.4.1]{thomas2006elements}.    

\subsubsection{Source-Channel Separation Theorem with Distortion} 
The communication process may be generally treated as that the encoded source passes through a noisy channel, where the distortion achieved at the receiving end depends on both the lossy source coding rate and the channel capacity. The \ac{sct} with distortion demonstrates the fact that a lossy source code with a rate $R(D)$, can be recovered with distortion $D$ after passing through a channel with capacity $C$, if and only if \cite{thomas2006elements}          
\begin{equation} \label{rdc}
R(D) < C.
\end{equation}     

\subsubsection{Quadratic Gaussian Problem}\label{prerw} 
In general, it is a challenging task to derive the explicit expression of the \ac{rd} function in \eqref{rd1} for an arbitrary source distribution and distance metric. For analysis convenience, we focus on the \emph{quadratic Gaussian problem} \cite{8744500,8636539}. Specifically, for independent Gaussian random variables $\mathsf{A}_i \sim \mathcal{CN}(0,\lambda_i), \kern 2pt i=1,2,\cdots,N$, let the distortion measure be the squared-error, i.e., $d(a^N,\hat{a}^N)=\sum_{i=1}^N (a_i-\hat{a}_i)^2$. Then the explicit expression of \ac{rd} function in \eqref{rd1} can be given by \cite[Th. 10.3.3]{thomas2006elements} \footnote{The coefficient $\frac{1}{2}$ is ignored since complex variables are considered.}        
\begin{equation}
R(D) = \sum_{i=1}^{N} \log \frac{\lambda_i}{D_i}. \kern 5pt 
\end{equation}
The distortion $D_i$ exhibits a reverse water-filling form as 
\begin{equation}\label{rw1}
	\sum_{i=1}^{N} D_i=D, \kern 5pt
	D_i = \left\{
	\begin{aligned}
		\xi, \kern 5pt \text{if} \kern 2pt \xi < \lambda_i, \\
		\lambda_i, \kern 5pt \text{if} \kern 2pt \xi \ge \lambda_i.
	\end{aligned} \right.
\end{equation}
where $\xi$ represents the reverse water-filling factor. The right equation in \eqref{rw1} can also be written by $D_i = \lambda_i - (\lambda_i-\xi)^+$. 

\subsection{Optimal Structures of \ac{snc} Waveforms} \label{presc} 

 
For decades, \ac{snc} researchers have been mostly working on the general Gaussian linear model as follows
\begin{equation}\label{glm1}
\mathbf{Y}=\mathbf{H}\mathbf{X}+\mathbf{Z},
\end{equation}  
where $\mathbf{Y} \in \mathbb{C}^{M \times T}$, $\mathbf{X}\in \mathbb{C}^{N \times T}$, and $\mathbf{Z}\in \mathbb{C}^{M \times T}$ are the received signal, transmitted signal, and additive Gaussian noise, respectively. $\mathbf{H}\in \mathbb{C}^{M \times N}$ represents the transmission channel that is known in the communication systems but to be estimated in the sensing systems. $T$, $N$, and $M$ are the numbers of symbols, transmitting and receiving antennas at the BS, respectively. We assume that the entries of $\mathbf{Z}$ follow the complex Gaussian distribution with $\mathcal{CN}(0,\sigma^2)$. 

The purpose of \ac{snc} waveform design is to find an appropriate transmission strategy $\mathbf{X}$ to optimize the performance based on some criteria while satisfying the power budget. In what follows, we will introduce two well-studied waveform design schemes and the structures of optimal solutions $\mathbf{X}^\star$ for \ac{snc} systems, which will be differentiated by the subscripts $s$ and $c$, respectively.  


\subsubsection{Communication- or \ac{mi}-Optimal Waveform} For communication systems, the transmitted waveform $\mathbf{X}_c$ is a random signal unknown at the receiver. With the Gaussian signal $\mathbf{X}_c \sim \mathcal{CN}(\mathbf{0},\mathbf{R}_c)$, the \ac{mi} between the received and transmitted signals conditioned on communication channel $\mathbf{H}_c$ can be characterized by
\begin{equation} \label{cap}
I(\mathbf{Y}_c;\mathbf{X}_c|\mathbf{H}_c) = 
 \logdet{\frac{1}{\sigma^2_c}\mathbf{H}_c\mathbf{R}_c\mathbf{H}_c^H+\mathbf{I}_N} \triangleq I_c(\mathbf{R}_c).
\end{equation}  
where $\mathbf{R}_c \in \mathbb{C}^{N \times N}$ is the statistical covariance matrix, and the channel $\mathbf{H}_c$ is assumed to be perfectly known. The \ac{mi}-optimal waveform design problem may be formulated as
\begin{equation}\label{cm}
\mathop {\max} \limits_{\mathbf{R}_c} \kern 2pt I_c(\mathbf{R}_c), \kern 5pt \text{s.t.} \kern 2pt \tr{\mathbf{R}_c} \le P_T,
\end{equation}
where $P_T$ is the power budget. The global optimal solution $\mathbf{R}_c^\star$ can be readily obtained since problem \eqref{cm} is convex \ac{wrt} the matrix variable. By denoting the eigenvalue decomposition of $\mathbf{R}_c^\star = \mathbf{\Psi}_c \bm{\Lambda}_c \mathbf{\Psi}_c^H$, then the optimal waveform $\mathbf{X}_c^\star$ may be reconstructed by
\begin{equation}\label{ccop} 
\mathbf{X}_c^\star = \mathbf{\Psi}_c \bm{\Lambda}_c \mathbf{Q}_c^H, 
\end{equation}
where $\mathbf{Q}_c$ contains \ac{iid} circularly symmetric complex entries satisfying $\mathept{\mathbf{Q}_c\mathbf{Q}_c^H}=\mathbf{I}_N$. 

Let us denote the positive semidefinite matrix $\bm{\Sigma}_c=\mathbf{H}_c^H\mathbf{H}_c$, where $\lambda_{c_i} \ge 0$ and $\mathbf{U}_c$ are the eigenvalues and the eigenspace of $\bm{\Sigma}_c$, respectively. Thus, it is well-known that the optimal solution \eqref{ccop} admits the following specific structure
\begin{equation} 
\mathbf{\Psi}_c = \mathbf{U}_c, \kern 2pt \bm{\Lambda}_c = \diag{\Big(\zeta_c-\frac{\sigma^2_c}{\lambda_{c_1}}\Big)^+ ,\cdots, \Big(\zeta_c-\frac{\sigma^2_c}{\lambda_{c_N}}\Big)^+ }^{\frac{1}{2}}, 
\end{equation}  
where $\zeta_c$ is the water-filling factor satisfying
\begin{equation}\label{waterfilling}
\sum_{i=1}^{N}\left(\zeta_c-\frac{\sigma^2_c}{\lambda_{c_i}}\right)^+ = P_T.
\end{equation}
Hereafter, we term the unitary matrix $\mathbf{U}_c$ as the \emph{communication subspace} based on this specific structure.

\subsubsection{Sensing- or \ac{mmse}-Optimal Waveform}\label{press}
The sensing task is to estimate $\mathbf{H}_s$, or, more relevant to radar sensing, to estimate the target parameters (e.g., amplitude, delay, angle, and Doppler) contained in $\mathbf{H}_s$ as accurate as possible. In this paper, we consider the waveform design for a general TRM estimation task. By vectorizing the Hermitian of the received signal in (\ref{glm1}), we have    
\begin{equation}\label{vecs}
\mathbf{y}_s=\tilde{\mathbf{X}}_s\mathbf{h}_s+\mathbf{z}_s,
\end{equation}  
where $\mathbf{y}_s=\text{vec}(\mathbf{Y}_s^H)$, $\tilde{\mathbf{X}}_s=\mathbf{I}_{M_s} \otimes \mathbf{X}_s^H$, $\mathbf{h}_s=\text{vec}(\mathbf{H}_s^H)$, $\mathbf{z}_s = \text{vec}(\mathbf{Z}_s^H)$. TRM estimation is to obtain the $\mathbf{h}_s$ from noisy received signal $\mathbf{y}_s$. 

\textit{\textbf{Assumption 1:}} The random vector $\mathbf{h}_s$ follows complex Gaussian distribution $\mathcal{CN}(0,\mathbf{I}_{M_s} \otimes \bm{\Sigma}_s)$ with $\bm{\Sigma}_s \in \mathbb{C}^{N \times N}$ being the covariance matrix of each column of $\mathbf{H}_s$.\footnote{We adopt this Kronecker structure for mathematical simplicity and ease of discussion. This assumption implies that the receiving antennas are sufficiently isolated, allowing us to ignore the correlations among the rows of $\mathbf{H}_s$ \cite{8579200}. No constraints are imposed on the transmitting array.} 

By adopting Assumption 1, the \ac{mmse} between $\mathbf{h}_s$ and its estimate $\hat{\mathbf{h}}_s$ may be expressed by  
\begin{equation}\label{mmse1}
\mathept{\|\mathbf{h}_s-\hat{\mathbf{h}}_s\|^2} = M_s \tr{ \Big( \frac{1}{\sigma^2_s} \mathbf{X}_s\mathbf{X}_s^H + \bm{\Sigma}_s^{-1} \Big)^{-1} }.
\end{equation}
Let $\mathbf{R}_s = 1/T \mathbf{X}_s\mathbf{X}_s^H$ be the sample covariance matrix. The \ac{mmse}-optimal waveform design may be formulated by \footnote{ The coefficient $T$ is absorbed into the power for notational consistency.} 
\begin{equation} \label{rmse}
\mathop {\min} \limits_{\mathbf{R}_s} \kern 2pt \tr{ \Big( \frac{1}{\sigma^2_s} \mathbf{R}_s + \bm{\Sigma}_s^{-1} \Big)^{-1} }, \kern 5pt \text{s.t.} \kern 2pt \tr{\mathbf{R}_s} \le P_T.
\end{equation}
\eqref{rmse} is a convex problem and the optimal solution $\mathbf{R}_s^\star$ can be readily obtained. Following the similar procedure in \eqref{ccop}, the optimal waveform $\mathbf{X}_s^\star$ may be reconstructed by 
\begin{equation} \label{ssop}
\mathbf{X}_s^\star = \mathbf{\Psi}_s \bm{\Lambda}_s \mathbf{Q}_s^H,   
\end{equation}
which also yields the following specific structure \cite{4194775}    
\begin{equation} \label{ops}
\mathbf{\Psi}_s = \mathbf{U}_s, \kern 2pt \bm{\Lambda}_s = \diag{\Big(\zeta_s-\frac{\sigma^2_s}{\lambda_{s_1}}\Big)^+ ,\cdots, \Big(\zeta_s-\frac{\sigma^2_s}{\lambda_{s_N}}\Big)^+ }^{\frac{1}{2}}, 
\end{equation}
where $\lambda_{s_i} \ge 0$ and $\mathbf{U}_s$ are the eigenvalues and the corresponding eigenspace of matrix $\bm{\Sigma}_s$, respectively. $\mathbf{Q}_s$ is an arbitrary semi-unitary matrix, and $\zeta_s$ represents the water filling factor similar to \eqref{waterfilling}. The detailed derivations can be found in \cite{4194775}. Also, we term the unitary matrix $\mathbf{U}_s$ as the \emph{sensing subspace}.  

\textbf{Remark 1:} The right singular spaces of \ac{snc} waveforms, namely, the random matrix $\mathbf{Q}_c$ and the semi-unitary matrix $\mathbf{Q}_s$, cannot be directly determined using the MI- and MMSE- optimal criteria in \eqref{cm} and \eqref{rmse}. In other words, these matrices are irrelevant to the \ac{snc} channel conditions $\bm{\Sigma_s}$ and $\bm{\Sigma_c}$. Although they can be further refined by imposing symbol-level constraints, such as constant modulus or similarity requirements \cite{6649991}, this is beyond the scope of this paper. As a result, we reformulate the expressions of \ac{mi} and \ac{mmse} as follows
\begin{subequations}
\begin{align}
I_c(\mathbf{\Psi}_c,\bm{\Lambda}_c) &= \logdet { \frac{1}{\sigma^2_c}\mathbf{\Psi}_c\bm{\Lambda}_c^2\mathbf{\Psi}_c^H\bm{\Sigma}_c^H+\mathbf{I}_N}, \label{ccs} \\
D_s(\mathbf{\Psi}_s,\bm{\Lambda}_s) &= M_s \tr { \Big( \frac{1}{\sigma^2_s} \mathbf{\Psi}_s\bm{\Lambda}_s^2\mathbf{\Psi}_s^H + \bm{\Sigma}_s^{-1} \Big)^{-1} }, \label{dds}
\end{align}  
\end{subequations}
Here, the variable $\mathbf{\Psi}$ acts as the beamforming design, and the variable $\bm{\Lambda}$ represents the power allocation.

\section{The CAS Framework}\label{SystemModel}
In this section, we establish the \ac{cas} framework based on the \ac{rd} and \ac{sct} theorems. The EC scheme is adopted where the BS transmits the estimates of the parameters rather than the raw data to the end-user. Then, we formulate a general waveform design problem within the \ac{cas} framework and derive the explicit expressions of the \ac{mmse} and \ac{rd} function in the scenario of TRM estimation.    
 
\subsection{System Model and Problem Formulation}
\begin{figure}[!t]
	\centering
	\includegraphics[width= 3 in]{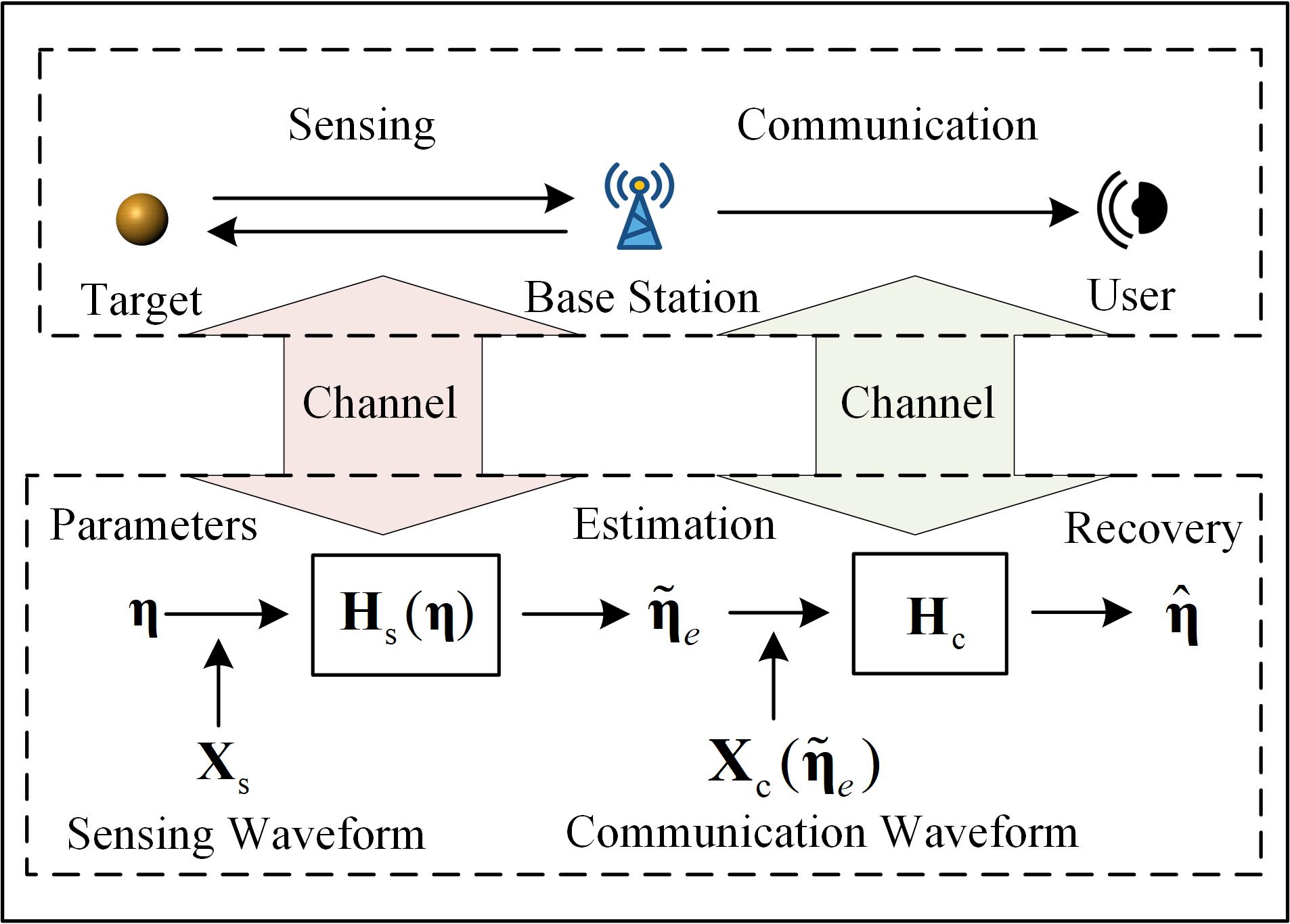}
	\caption{The to-be-estimated parameters successively ``pass through'' the \ac{snc} channels to the user.}
	\label{passthrough}
\end{figure}

Let us consider a BS equipped with $N$ transmitting and $M_s$ receiving antennas. The BS actively performs wireless sensing tasks, such as area monitoring and object recognition, and then delivers the sensed data to an $M_c$-antenna user. Let $\bm{\upeta} \in \mathbb{R}^K$ denote the parameter vector to be sensed, which may represent the target state, environmental features, or other relevant variables, taking values in a set $\mathcal{A}$ with a prior distribution $p_{\bm{\upeta}}(\bm{\eta})$. As shown in Fig. \ref{passthrough}, the \ac{cas} process can be equivalently viewed as the target information sequentially ``passes through'' the \ac{snc} channels before reaching the user. In this scenario, the original parameter $\bm{\upeta}$, the estimated $\tilde{\bm{\upeta}}_e$ at the BS, and the recovered $\hat{\bm{\upeta}}$ at the user forms a Markov chain $\bm{\upeta} \to \tilde{\bm{\upeta}}_e \to \hat{\bm{\upeta}}$. The detailed signal models are provided below.   

$\bullet$ \textbf{Sensing Process (SP)}: The BS transmits sensing waveform to the target and yields the estimated parameter, namely, $\tilde{\bm{\upeta}}_e$ through the noisy echo signals. Similar to (\ref{glm1}), the received signal model can be expressed by
\begin{equation}\label{SensingM}
	\mathbf{Y}_s=\mathbf{H}_s(\bm{\upeta})\mathbf{X}_s+\mathbf{Z}_s,
\end{equation} 
where $\mathbf{H}_s(\bm{\upeta})$ represents the sensing channel matrix \ac{wrt} the latent parameters $\bm{\upeta}$. 

$\bullet$ \textbf{Communication Process (CP)}: The estimated information $\tilde{\bm{\upeta}}_e$ is transmitted to the user, then be reconstructed as an estimate $\hat{\bm{\upeta}}$ from the following received signal  

\begin{equation}
	\mathbf{Y}_c=\mathbf{H}_c\mathbf{X}_c(\tilde{\bm{\upeta}}_e)+\mathbf{Z}_c.
\end{equation} 
Here, we use $\mathbf{X}_c(\tilde{\bm{\upeta}}_e)$ to emphasize the waveform carrying the information of $\tilde{\bm{\upeta}}_e$ through source-channel coding. 

By adopting \ac{mse} as the distance metric, the \ac{snc} distortions can be respectively defined by
\begin{equation}
 D_s = \mathept{\left\|\bm{\upeta}-\tilde{\bm{\upeta}}_e\right\|_2^2}, \kern 5pt D_c = \mathept{\left\|\tilde{\bm{\upeta}}_e-\hat{\bm{\upeta}}\right\|_2^2}. 
\end{equation}
In the proposed CAS framework, we really concern about the CAS distortion defined by the average distortion between the recovery $\hat{\bm{\upeta}}$ and the ground truth $\bm{\upeta}$, i.e.,      
\begin{equation} {\label{Dsc}} 
D=\mathept{\left\|\bm{\upeta}-\hat{\bm{\upeta}}\right\|_2^2}.
\end{equation}
which is influenced by both the sensing distortion $D_s$ in the SP and the communication distortion $D_c$ in the CP. Additionally, the sensing distortion $D_s$ is directly determined by the sensing waveform $\mathbf{X}_s$, while the communication distortion $D_c$ is bounded by the achievable rate $I_c$ which depends on the communication waveform $\mathbf{X}_c$, as discussed in Section \ref{preloss}. This interdependence motives us to seek for an optimal waveform that minimizes the CAS distortion. The general optimization problem can be formulated as \footnote{Here, we slightly modify the \ac{sct} constraint by transforming the strict "$<$" of \eqref{rdc} into "$\le$". On one hand, this modification can significantly simplify the analysis and solution of the optimization problem. On the other hand, different from the information theory, the error caused by this operation is tolerable in practical applications.}  
\begin{equation} \label{p0}
\mathcal{P}_0 \left\{	\begin{aligned}
\mathop { \text{min} }\limits_{\mathbf{X}} \kern 5pt & D   \\
\text{s.t.} \kern 7pt & R(D_c) \le I_c, \kern 2pt \mathept{b(\mathbf{X})} \le B.  
\end{aligned} \right.
\end{equation}
The first constraint is SCT to guarantee reliable sensory data transmission, while the second constraint is related to system resource, including power, bandwidth, etc. The function $b(\mathbf{X})$ denotes the cost function with $B$ being the corresponding resource budget. Note that deriving explicit expressions for problem $\mathcal{P}_0$ \ac{wrt} the variable $\mathbf{X}$ is generally a challenging task. In this paper, we focus on the quadratic Gaussian problem for TRM estimation, i.e., $\bm{\upeta}=\mathbf{h}_s$, as demonstrated in Section \ref{prelimi}. Beyond optimal waveform design, we aim to reveal the performance tradeoffs between the SP and CP, thereby providing insights into the \ac{cas} framework. More detailed sensing tasks are left for our future research.
         
\subsection{Target Response Matrix Estimation}
Recall that the received signal \eqref{vecs} becomes Gaussian linear model under the Assumption 1. Hence, the \ac{mmse} estimator is recognized as the optimal linear estimator for TRM estimation in the SP. In what follows, we elucidate the relationship between CAS distortion and \ac{snc} distortions. Subsequently, we derive the explicit expressions for the \ac{mmse} and \ac{rd} function in problem $\mathcal{P}_0$.

\subsubsection{Objective Function}
While adopting the \ac{mmse} estimator in the SP, i.e., $\tilde{\bm{\upeta}}_e = \mathbb{E}\left[ \bm{\upeta}|\mathbf{y}_s\right]$, the CAS distortion for TRM estimation can be equivalently decomposed into the sum of the $D_s$ and $D_c$ \footnote{ It is important to note that the equality \eqref{Dsc2} does not necessarily hold for arbitrary sensing tasks which cannot benefit from the conditional expectation.}, i.e.,  
\begin{equation} {\label{Dsc2}} 
D=\mathbb{E}\left[\left\|\bm{\upeta}-\tilde{\bm{\upeta}}_e+\tilde{\bm{\upeta}}_e-\hat{\bm{\upeta}}\right\|_2^2\right] \mathop = \limits^{(a)} D_s+D_c,
\end{equation}     
where $(a)$ holds due to the conditional expectation
\begin{equation}
\mathbb{E}\big[(\bm{\upeta}-\tilde{\bm{\upeta}}_e)^T(\tilde{\bm{\upeta}}_e-\hat{\bm{\upeta}})\big]= \mathbb{E}\big[(\bm{\upeta}-\mathbb{E}\left[ \bm{\upeta}|\mathbf{y}_s\right])^T\phi(\mathbf{y}_s)\big]=0. \nonumber
\end{equation}
Here, $\phi(\mathbf{y}_s)$ represents an arbitrary function \ac{wrt} $\mathbf{y}_s$ \cite{8636539}. 

\subsubsection{Estimate $\tilde{\bm{\upeta}}_e$ and Distortion $D_s$}
The sensing distortion is equivalent to the achieved MMSE, i.e., $D_s = \text{MMSE}$, whose expression is given in \eqref{dds}. In this context, the estimate $\tilde{\bm{\upeta}}_e$ can be expressed by \cite{BookEstimationTheory} 
\begin{equation} \label{estimate111}
\tilde{\bm{\upeta}}_e = \Big(\mathbf{I}_{M_s} \otimes \big(\bm{\Sigma}_s\mathbf{X}_s\mathbf{R}_y^{-1}\big)\Big)\textbf{y}_s,
\end{equation}  
where the observations $\mathbf{y}_s$ follow a Gaussian distribution $\mathbf{y}_s \sim \mathcal{CN}(\mathbf{0},\mathbf{I}_{M_s} \otimes \mathbf{R}_y)$ with $\mathbf{R}_y$ defined by
\begin{equation}
\mathbf{R}_y \triangleq \mathbf{X}_s^H \bm{\Sigma}_s \mathbf{X}_s + \sigma^2_s\mathbf{I}_T.
\end{equation}  
It is evident that the estimate $\tilde{\bm{\upeta}}_e$ in \eqref{estimate111} follows a complex Gaussian distribution $ \mathcal{CN}(\mathbf{0},\mathbf{I}_{M_s} \otimes \mathbf{R}_{\upeta})$, where $\mathbf{R}_{\upeta}$ is given by \footnote{At a glance, the covariance of the estimate corresponds to the difference between the original parameter covariance and the reduction in covariance achieved by MMSE estimation. This implies that a lower MMSE results in higher residual covariance, which, in turn, requires greater communication channel capacity to transmit the estimate, leading to an explicit performance tradeoff between \ac{sp} and \ac{cp}. This relationship will be further illustrated in Section \ref{discussion} and Fig. \ref{New_Waterfill}. } 
\begin{equation}\label{relationrs}
\mathbf{R}_{\upeta} = \bm{\Sigma}_s - \Big( \frac{1}{\sigma^2_s} \mathbf{X}_s\mathbf{X}_s^H + \bm{\Sigma}_s^{-1} \Big)^{-1}.
\end{equation}    
Applying the waveform structure \eqref{ssop}, the matrix $\mathbf{R}_{\upeta}$ can be recast by 
\begin{equation}\label{reta}
\mathbf{R}_{\upeta} = \bm{\Sigma}_s\mathbf{\Psi}_s\bm{\Lambda}_s\Big(\bm{\Lambda}_s\mathbf{\Psi}_s^H\bm{\Sigma}_s\mathbf{\Psi}_s\bm{\Lambda}_s+\sigma^2_s\mathbf{I}_T\Big)^{-1}\bm{\Lambda}_s\mathbf{\Psi}_s^H\bm{\Sigma}_s^H, 
\end{equation}
which is related to $\mathbf{\Psi}_s$ and $\bm{\Lambda}_s$.

\subsubsection{Recovery Distortion $D_c$}
In the CP, the Gaussian random variables $\tilde{\bm{\upeta}}_e$ is transmitted to the user through source-channel coding. Note that unitary rotation would not cause any information loss in Gaussian variables. Let us denote $\tilde{\bm{\upeta}}_s = \mathbf{U}_{\upeta}^H\tilde{\bm{\upeta}}_e$, where $\mathbf{U}_{\upeta}$ is the eigenspace of matrix $\mathbf{R}_{\upeta}$. Thus, we have 
\begin{equation}
\tilde{\bm{\upeta}}_s \sim \mathcal{CN}(\mathbf{0},\mathbf{I}_{M_s} \otimes \diag{\lambda_1(\mathbf{R}_{\upeta}), \cdots, \lambda_N(\mathbf{R}_{\upeta})}),
\end{equation}        
where $\lambda_i(\mathbf{R}_{\upeta})$ represents the $i$-th eigenvalue of matrix $\mathbf{R}_{\upeta}$. Based on the results in Section \ref{prerw}, the \ac{rd} function can be obtained by
\begin{equation} \label{rdcc}
R(D_c) = M_s\sum_{i=1}^{N} \log \frac{\lambda_i(\mathbf{R}_{\upeta})}{D_{c_i}}, 
\end{equation}     
which is in a reverse water-filling form as
\begin{equation} \label{Dcr}
D_c = M_s\sum_{i=1}^{N}D_{c_i}=M_s\sum_{i=1}^{N} \lambda_i(\mathbf{R}_{\upeta})-\big(\lambda_i(\mathbf{R}_{\upeta})-\xi\big)^+.
\end{equation}

As the SP and CP are performed on a single hardware platform, we consider the power constraint with $P_T$ being the power budget. Consequently, with the achievable rate $I_c(\mathbf{\Psi}_c,\bm{\Lambda}_c)$ in \eqref{ccs}, the original problem $\mathcal{P}_0$ can be transformed into the following form
\begin{equation} \label{p1}
	\mathcal{P}_1 \left\{	\begin{aligned}
		\mathop { \text{min} }\limits_{\mathbf{\Psi}_c, \mathbf{\Psi}_s, \bm{\Lambda}_c, \bm{\Lambda}_s} & \kern 3pt  D_s + D_c   \\
		\text{s.t.} \kern 18pt & M_s\sum_{i=1}^{N} \log \frac{\lambda_i(\mathbf{R}_{\upeta})}{D_{c_i}} \le I_c(\mathbf{\Psi}_c,\bm{\Lambda}_c), \\
		& \eqref{reta}, \kern 5pt \eqref{Dcr}, \kern 5pt \text{Tr}(\bm{\Lambda}_c^2)+\text{Tr}(\bm{\Lambda}_s^2) \le P_T.
	\end{aligned} \right.
\end{equation}
Apparently, addressing the non-convex problem $\mathcal{P}_1$ presents inherent challenges due to two primary reasons: 1) The eigenvalue $\lambda_i(\mathbf{R}_{\upeta})$ of matrix $\mathbf{R}_{\upeta}$ is still an implicit function \ac{wrt} the variables $\mathbf{\Psi}_s$ and $\bm{\Lambda}_s$. 2) The imposed reverse water-filling constraint introduces an unknown nuisance parameter (i.e., the factor $\xi$), which may only be determined through a numerical search algorithm in general. In what follows, we will propose waveform design approaches for two typical signaling schemes within the ISAC system, i.e., the \ac{sw} and \ac{dw} designs. 
        
\section{Separated \ac{snc} Waveform Design}\label{sncwd}
In the \ac{sw} strategy, the BS transmits individual waveforms in the SP and CP by the shared use of the transmission power. This implies that the eigenspaces of \ac{sw} can be independently designed but the diagonal power allocation matrices are coupled through the power constraint. Let us denote
\begin{equation} \label{opstru}
\bm{\Lambda}_s = \diag{\mathbf{p}_s}^{\frac{1}{2}}, \kern 5pt
\bm{\Lambda}_c = \diag{\mathbf{p}_c}^{\frac{1}{2}},
\end{equation}   
with $\mathbf{p}_s=[p_{s_1},\cdots, p_{s_N}]^T$ and $\mathbf{p}_c=[p_{c_1},\cdots, p_{c_N}]^T$ being the power allocation vectors. Recalling the optimal waveform structures in Section \ref{presc}, we immediately have $\mathbf{\Psi}_s=\mathbf{U}_s$ and $\mathbf{\Psi}_c=\mathbf{U}_c$. Thus, by substituting the subspace structures, the corresponding expressions in problem $\mathcal{P}_0$ can be further simplified to 
\begin{subequations}
\begin{align}
& I_c(\mathbf{p}_c) = \sum_{i=1}^{N}\log \left(\frac{\lambda_{c_i}}{\sigma_c^2}p_{c_i}+1\right), \label{CC} \\ 
& D_s(\mathbf{p}_s) = M_s\sum_{i=1}^{N} f_i(p_{s_i}) \triangleq M_s\sum_{i=1}^{N}\frac{\lambda_{s_i}\sigma^2_s}{\sigma^2_s+\lambda_{s_i}p_{s_i}}, \label{DS} \\
& \lambda_i(\mathbf{R}_{\upeta}) = g_i(p_{s_i}) \triangleq  \frac{\lambda_{s_i}^2p_{s_i}}{\sigma^2_s+\lambda_{s_i}p_{s_i}}. \label{dd26}
\end{align}
\end{subequations}
From \eqref{dd26}, we can observe that the eigenvalue $\lambda_i(\mathbf{R}_{\upeta})$ can be explicitly expressed by the function of the variable $p_{s_i}$ thanks to the optimal subspace structure, thereby circumventing the eigenvalue decomposition. Moreover, the relationship in \eqref{relationrs} is clearly shown in the scalar form, i.e., $g_i(p_{s_i}) = \lambda_{s_i} - f_i(p_{s_i}) $, where the functions $g_i$ and $f_i$ are defined in \eqref{DS} and \eqref{dd26}, respectively. Consequently, the original problem $\mathcal{P}_1$ with $N \times N$-dimension matrix variables can be simplified into $N$-dimension vector variables, given by  
\begin{equation} \label{p2}
	\mathcal{P}_2 \left\{	\begin{aligned}
		\mathop { \text{min} }\limits_{\mathbf{p}_s, \mathbf{p}_c} \kern 3pt &  M_s\sum_{i=1}^{N} \left(\lambda_{s_i}-\big(g_i(p_{s_i})-\xi\big)^+\right)   \\
		\text{s.t} \kern 5pt &  M_s \sum_{i=1}^{N} \log \frac{g_i(p_{s_i})}{g_i(p_{s_i})-\big(g_i(p_{s_i})-\xi\big)^+} \le I_c(\mathbf{p}_c),\\
		& \mathbf{1}_N^T\big(\mathbf{p}_s+\mathbf{p}_c\big) \le P_T, \kern 5pt \mathbf{p}_s, \mathbf{p}_c \ge 0. 
	\end{aligned} \right.
\end{equation}          
Problem $\mathcal{P}_2$ is exactly a power allocation problem, which aims at minimizing the CAS distortion under the constraints of \ac{sct} and power budget. Before solving $\mathcal{P}_2$, we first show that both constraints are active at the optimal solution.    

\noindent
\underline{\textbf{\textit{Proposition 1:}}} The optimal solutions $\mathbf{p}_s^\star$ and $\mathbf{p}_c^\star$ are achieved if and only if the equality holds in both the \ac{sct} and power constraints.

\noindent   
\textbf{\textit{Proof:}} Please see Appendix \ref{apA}. $\hfill\blacksquare$  

\subsection{A General One-Dimensional Search Algorithm }

Let us denote $P_s$ and $P_c$ as the power allocated to the SP and CP, respectively, satisfying $P_s+P_c=P_T$. According to Proposition 1, we will show that the solution $(\mathbf{p}_s^\star,\mathbf{p}_c^\star)$ is uniquely determined once the power allocation pair $(P_s,P_c)$ is given. This implies that we can just search the optimal power allocation pair $(P_s,P_c)$ instead of solving problem $\mathcal{P}_2$ directly.        
         
\subsubsection{Maximizing Achievable Rate} The power $P_c$ allocated to the CP only affects on the achievable rate $I_c(\mathbf{p}_c)$ in $\mathcal{P}_2$. As previously discussed, the optimal solution $\mathbf{p}_c^\star$ of maximizing the achievable rate can be expressed by
\begin{equation} \label{copt}
p_{c_i}=\Big(\zeta_c-\frac{\sigma^2_c}{\lambda_{c_i}}\Big)^+, \kern 5pt \sum_{i=1}^{N}p_{c_i}=P_c. 
\end{equation}      

\subsubsection{Minimizing Sensing Distortion} Similarly, for a given power $P_s$, the minimum sensing distortion can be attained by the following solution     
\begin{equation}\label{sopt}
	p_{s_i}=\Big(\zeta_s-\frac{\sigma^2_s}{\lambda_{s_i}}\Big)^+, \kern 5pt \sum_{i=1}^{N}p_{s_i}=P_s. 
\end{equation}

\begin{algorithm}[!t]
\caption{One-Dimensional Search Algorithm}
\label{alg1}
\textbf{Initialize:} Input $L$, $P_T$, $\epsilon>0$, $p_{\min} = 0, \kern 2pt p_{\max} = P_T$. \\
\While{$\big|D_{k+1}-D_{k}\big|>\epsilon$}{
\For{$l=0:L$}{
$P_{s}^{(l)} =p_{\min}+\frac{l}{L}(p_{\max}-p_{\min}), P_{c}^{(l)} = P_T-P_{s}^{(l)}$\; 
Obtain $\mathbf{p}^{(l)}_{c}$ by water-filling algorithm in \eqref{copt} \; 
Calculate achievable rate $I_c(\mathbf{p}^{(l)}_{c})$ in \eqref{CC}\;
Obtain $\mathbf{p}_{s}^{(l)}$ by water-filling algorithm in \eqref{sopt}\;
Calculate $D_{s}^{(l)}(\mathbf{p}_{s}^{(l)})$ and $g_{i}^{(l)}$ in \eqref{DS}, \eqref{dd26} \;
Calculate $D_{c}^{(l)}$ by reverse water-filling in \eqref{rw1}\;
$\mathcal{D} \gets D^{(l)}=D_{s}^{(l)}+D_{c}^{(l)}$ \;
}
$D^{(l^*)} = \min D \in \mathcal{D}$  \;
$p_{\min} \gets P_s^{(l^*-1)}, p_{\max} \gets P_s^{(l^*+1)}, D_k \gets D^{(l^*)}$ \;
}
\KwOut{$(\mathbf{p}_s^\star,\mathbf{p}_c^\star)$ for $D^\star $.}
\end{algorithm}

\subsubsection{Calculating Communication Distortion} With the power allocation $\mathbf{p}_s^\star$ in \eqref{sopt} at hand, we can immediately compute the source variance $g_i$ through \eqref{dd26}. According to proposition 1, the maximum available coding rate is achieved by $R=I_c(\mathbf{p}_c)$. For a given rate and source variance, the communication distortion can be computed by the reverse water-filling algorithm in \eqref{rw1}. Thus, the CAS distortion is equal to $D_s+D_c$.       

Following the above steps, we may identify the optimal solution $(\mathbf{p}_s^\star,\mathbf{p}_c^\star)$ and the CAS distortion $D$ with a given power allocation pair $(P_s,P_c)$. The remaining work is to find optimal $(P_s^\star,P_c^\star)$ that minimizes $D$ over all the possible pairs. Consequently, the $N$-dimensional vector optimization problem can be further reduced to a one-dimensional search problem. Specifically, let $P_s$ take values over the interval $[0,P_T]$ with $L$ grids and $P_c = P_T-P_s$. The minimum $D$ corresponds to the optimal power allocation vectors $(\mathbf{p}_s^\star,\mathbf{p}_c^\star)$. The detailed algorithm flow is summarized in Algorithm \ref{alg1}.

\subsection{Special Case: I.I.D. Sensing Subchannels }
To further reveal the tradeoff between the SP and CP, we focus on a special case that the entries of TRM are \ac{iid} white Gaussian distributed, namely,

\textbf{\textit{Assumption 2:}} The vector $\mathbf{h}_s$ follows $\mathcal{CN}(0,\lambda_s\mathbf{I}_{M_sN})$. 

In this scenario, the optimal power allocation vector $\mathbf{p}_s$ in \eqref{sopt} for a given $P_s$ is the uniform allocation, where $p_{s_1} = \cdots = p_{s_N} = P_s/N$ since each subchannel experiences identical conditions. By substituting $\mathbf{p}_s$ into \eqref{DS} and \eqref{dd26}, we have   
\begin{equation}
\begin{aligned}
f_1(p_{s_1}) = \cdots = f_N(p_{s_N}) = \frac{N\lambda_s\sigma^2_s}{N\sigma^2_s+\lambda_sP_s} \triangleq f(P_s),  \\
g_1(p_{s_1}) = \cdots = g_N(p_{s_N}) = \frac{\lambda_s^2P_s}{N\sigma^2_s+\lambda_sP_s} \triangleq g(P_s).
\end{aligned}
\end{equation} 
Accordingly, $g(P_s) - \xi > 0$ is necessary for a positive achievable rate $R = I_c(\mathbf{p}_c)$ \footnote{Otherwise, the coding rate $R$ is equal to zero in terms of the reverse water-filling algorithm.}. The communication distortion is then given by $D_{c_i}=g(P_s)-(g(P_s) - \xi)^+=\xi$. By recalling proposition 1, the \ac{sct} constraint must satisfy 
\begin{equation}\label{34}
M_sN\log\frac{g(P_s)}{\xi} = I_c(P_T-P_s).
\end{equation}
Fortunately, we can express the reverse water-filling factor $\xi$ as a function of $P_s$, circumventing the need for numerical searching. By substituting $\xi$ into the objective, problem $\mathcal{P}_2$ can be reformulated as an unconstrained optimization problem \ac{wrt} the scalar variable $P_s$, namely,                    
\begin{equation}
\mathcal{P}_3: \mathop { \min }\limits_{P_s}  \kern 2pt   NM_s\Big[\big(1-e^{-\frac{I_c(P_T-P_s)}{M_sN}}\big)f(P_s)+\lambda_se^{-\frac{I_c(P_T-P_s)}{M_sN}}\Big].
\end{equation}
\noindent
\underline{\textbf{\textit{Proposition 2:}}} $\mathcal{P}_3$ is a convex optimization problem.

\noindent   
\textbf{\textit{Proof:}} Please see Appendix \ref{apB}. $\hfill\blacksquare$ 

\textbf{Remark 2:} Problem $\mathcal{P}_3$ clearly shows the performance tradeoff between the SP and CP. Let us denote the objective of $\mathcal{P}_3$ as $h(x)$. Then it holds immediately that $h(0)=h(P_T)=\lambda_s$. The physical meanings involve that no useful information will be attained by the user if the total power is allocated to either SP or CP. In this case, the user may only ``guess'' the target parameters with the prior knowledge of variance $\lambda_s$. Considering the convexity of $\mathcal{P}_3$, there exists an optimal power allocation scheme between $[0,P_T]$. This exhibits a \emph{power allocation tradeoff} between \ac{snc} processes in the \ac{cas} systems, where extremely allocating resources to each side will lead to significant performance loss.  

\section{Dual-functional Waveform Design}\label{dualfuncwd}
In the previous section, the \ac{snc} processes utilize separate waveforms coupled only through transmission power. In this section, we consider that the BS transmits a dual-functional waveform $\mathbf{X}$, which simultaneously performs sensing tasks and conveys the data acquired in the last epoch to the end-user \footnote{ Considering the waveform structure $\mathbf{X}=\mathbf{\Psi}\bm{\Lambda}\mathbf{Q}^H$, the matrix $\mathbf{Q}$ should be random to convey useful information in the context of DW scheme. System design that accounts for the randomness of $\mathbf{Q}$ is beyond the scope of this paper. The readers are referred to \cite{10147248,lu2023random} for more details.}. Similar to problem $\mathcal{P}_1$, the design of the DW can be formulated by 
\begin{equation} \label{p4}
	\mathcal{P}_4 \left\{	\begin{aligned}
		\mathop { \text{min} }\limits_{\mathbf{\Psi}, \bm{\Lambda}} \kern 5pt & D_s + D_c   \\
		\text{s.t.} \kern 5pt & M_s\sum_{i=1}^{N} \log \frac{\lambda_i(\mathbf{R}_{\upeta})}{D_{c_i}} \le I_c(\mathbf{\Psi}, \bm{\Lambda}), \\
		&\eqref{reta}, \kern 5pt \eqref{Dcr}, \kern 5pt \text{Tr}(\bm{\Lambda}^2)\le P_T.
	\end{aligned} \right.
\end{equation}
In addition to the challenges encountered in solving problem $\mathcal{P}_3$, the variables $\mathbf{\Psi}, \bm{\Lambda}$ are coupled in both the power and \ac{sct} constraints of problem $\mathcal{P}_4$. This interdependence introduces the following two kinds of tradeoffs. 
  
\begin{itemize}
	\item \emph{Subspace tradeoff:} The left singular subspace $\mathbf{\Psi}$ may not align simultaneously with both the sensing subspace $\mathbf{U}_s$ and the communication subspace $\mathbf{U}_c$.  
	\item \emph{Water-filling tradeoff:} Although the total transmit power is shared between the SP and CP, the power allocation may not simultaneously satisfy the water-filling criteria for both the sensing subchannel $\sigma_s^2/\lambda_{s_i}$ and communication subchannel $\sigma_c^2/\lambda_{c_i}$.     
\end{itemize}
Unlike the SW scheme, the waveform design in $\mathcal{P}_4$ cannot be reduced to a low-dimensional power allocation problem due to the presence of the \textit{subspace tradeoff}, which prevents the direct determination of optimal subspace structure. Additionally, constructing a \ac{dw} that simultaneously optimizes both SP and CP is a very challenging task, unless the \ac{snc} channels are sufficiently similar. In this section, we develop a heuristic yet effective algorithm to address the non-convex problem $\mathcal{P}_4$ and a modified gradient projection algorithm for a special case with independent sensing subchannels.

\subsection{A Heuristic MI Maximization Algorithm}

\begin{algorithm}[!t]
	\caption{Heuristic MI maximization algorithm}
	\label{alg2}
	\textbf{Initialize:} The grid number $L$. \\
	\For{$l=0:L$}{
		$\alpha^{(l)}=l/L$\;
		Solve problem \eqref{relaxc} with $\alpha^{(l)}$ and obtain $\mathbf{R}^{(l)}$  \;
		Obtain $\mathbf{\Psi}^{(l)}, \bm{\Lambda}^{(l)}$ according to $\mathbf{R}^{(l)}$ \;
		Compute $D_s(\mathbf{R}^{(l)})$ according to \eqref{dds} \;
		Compute $I_c(\mathbf{R}^{(l)})$ according to \eqref{cap}\;     
		Compute $D_c(\mathbf{\Psi}^{(l)}, \bm{\Lambda}^{(l)})$ by reverse water-filling \;
		 $\mathcal{D} \gets D^{(l)}=D_s^{(l)}+D_c^{(l)}$\;
	}
	\KwOut{$\mathbf{R}^\star = \mathop { \arg \min } \limits_{D \in \mathcal{D}} \kern 2pt  D(\mathbf{R}) $.}
\end{algorithm} 

The primary difficulty in $\mathcal{P}_4$ arises from the SCT constraint, which involves transforming the estimation quantity (distortion) into an information quantity (rate) through the RD function. In \cite{7279172}, the radar estimation rate is defined for scalar time-delay estimation tasks, bridging the quantities between estimation and information in ISAC systems. Recently, this concept has been extended to vector cases within the quadratic Gaussian problem based on the \ac{rd} theory \cite{10192083}. These developments motivate us to adopt an MI maximization approach instead of directly minimizing distortion, thereby circumventing the complexities introduced by the SCT constraint.     
	  
Specifically, the sensing MI between observations and the TRM conditioned on the waveform $\mathbf{X}$ can be defined by \cite{4194775}   
\begin{equation}
I(\mathbf{Y}_s;\mathbf{H}_s|\mathbf{X}) = M_s \logdet{\frac{1}{\sigma^2_s} \mathbf{X}^H\bm{\Sigma}_s\mathbf{X}+\mathbf{I}_{T}}.  
\end{equation} 
Denoting the signal covariance matrix as $\mathbf{R} = 1/T\mathbf{X}\mathbf{X}^H$, the sensing MI can be recast as a function of $\mathbf{R}$, i.e., 
\begin{equation}
I_s(\mathbf{R}) = M_s \logdet{  \frac{1}{\sigma^2_s} \bm{\Sigma}_s\mathbf{R}+\mathbf{I}_{N}}.  
\end{equation} 
By choosing the reverse water-filling factor $\xi_s = \sigma_s^2/\zeta_s$ appropriately within the RD function, we have the following equality for Gaussian linear sensing model \eqref{SensingM}    
\begin{equation}\label{rdmmse}
R(D_s)=R(\text{MMSE})=I_s(\mathbf{R}).
\end{equation}
Equality \eqref{rdmmse} demonstrates that minimizing sensing distortion is equivalent to maximizing sensing MI. Note that this equivalence pertains not only to the identical solutions, as discussed in \cite{4194775}, but also to the one-one correspondence between the objectives. We refer the reader to \cite{10192083} for details. 

Similarly, the relationship between communication distortion and achievable rate is given by \footnote{We adopt the fact that the statistical covariance is approximately equal to the sample covariance for a large $T$.}
\begin{equation}\label{rdmmse2}
R(D_c) = I_c(\mathbf{R}) = \logdet{\frac{1}{\sigma^2_c}\mathbf{H}_c\mathbf{R}\mathbf{H}_c^H+\mathbf{I}_N},
\end{equation}
with an appropriately chosen the reverse water-filling factor $\xi_c$. Unfortunately, while this equivalence holds individually for both \ac{snc} processes, minimizing the sum distortion $D_s+D_c$ may not be equivalent to maximizing the sum MI $I_s(\mathbf{R})+I_c(\mathbf{R})$. This is due to the difficulty in finding a single factor $\xi = \xi_s = \xi_c $ that satisfies both the equalities \eqref{rdmmse} and \eqref{rdmmse2} simultaneously.\footnote{This is the reason why we term this MI maximization approach as ``Heuristic Algorithm''. However, simulation results demonstrate that this algorithm exhibit a outstanding performance.}. To tackle this problem, we introduce a to-be-determined weighting factor $\alpha$ to balance the \ac{snc} performance. Then, the weighted \ac{mi} maximization problem can be formulated by                          
\begin{equation} \label{relaxc}
	\begin{aligned}
		\mathop { \text{max} }\limits_{\mathbf{R}} \kern 3pt &  \alpha I_s(\mathbf{R}) + (1-\alpha)I_c(\mathbf{R})   \\
		\text{s.t.} \kern 5pt &\text{Tr}(\mathbf{R})\le P_T.
	\end{aligned}
\end{equation}
Problem \eqref{relaxc} is convex since both MIs are concave \ac{wrt} $\mathbf{R}$. The optimal waveform $\mathbf{X}$ may be immediately obtained by following a similar process to that in Section \ref{prelimi}. We adopt a search procedure over the interval $[0,1]$ to seek for the optimal $\alpha^\star$. Concretely, for a given $\alpha \in [0,1]$, we can obtain the solution of \eqref{relaxc} and then compute the CAS distortion. The optimal waveform corresponds to the minimum CAS distortion across all $\alpha$ values. The detailed algorithm flow is summarized in Algorithm \ref{alg2}.        

\textbf{Remark 3:} (1) Notably, the SCT constraint is omitted in the MI maximization problem \eqref{relaxc}. The SCT constraint can be viewed as a transformation from a distortion metric to an information metric. When focusing on the MI maximization problem to obtain the optimal solution $\mathbf{R}^\star$, all intermediate variables are determined accordingly. Therefore, the SCT constraint can always be satisfied by appropriately adjusting the reverse water-filling factor to minimize the CAS distortion. (2) The problem \eqref{relaxc} is similar to the MI-based waveform design framework discussed in \cite{10129042}. However, it is important to highlight that the objectives and motivations behind the problem formulation differ fundamentally. Our approach explicitly aims to minimize the CAS distortion by solving a series of optimization problems with varying values of $\alpha$. In contrast, most existing literature focuses solely on maximizing the sum of \ac{snc} MI, whether weighted or not, often without a clear physical interpretation.

\subsection{Special Case: Independent Sensing Subchannels}

In this subsection, we introduce the following assumption.

\textbf{\textit{Assumption 3:}} The covariance matrix $\mathbf{I}_{M_s} \otimes \bm{\Sigma}_s $ for the vector $\mathbf{h}_s$ is diagonal, and can be expressed as $\bm{\Sigma}_s = \bm{\Lambda}_{h_s} = \diag{\lambda_{s_1},\lambda_{s_2},\cdots,\lambda_{s_N}}$.

Under Assumption 3, the subspace tradeoff is eliminated because the sensing subspace $\mathbf{U}_s$ can be any arbitrary unitary matrix. This allows us to align the left singular subspace $\mathbf{\Psi}$ with the communication subspace $\mathbf{U}_c$. By denoting $\bm{\Lambda} = \diag{\mathbf{p}}$ and $\mathbf{p}=[p_1,p_2,\cdots, p_{N}]^T$, the problem $\mathcal{P}_4$ can be transformed into the following form       
\begin{equation}  
\mathcal{P}_5 \left\{ \begin{aligned}
\mathop { \text{min} }\limits_{\mathbf{p}} \kern 5pt & M_s\sum_{i=1}^{N} \Big(\lambda_s - \big(g_i(\mathbf{p})-\xi\big)^+ \Big)  \\
\text{s.t} \kern 5pt &  M_s \sum_{i=1}^{N} \log \frac{g_i(\mathbf{p})}{g_i(\mathbf{p}) - \big(g_i(\mathbf{p})-\xi\big)^+} \le I_c(\mathbf{p}),\\
& \mathbf{1}_N^T\mathbf{p} \le P_T, \kern 5pt \mathbf{p}\ge 0. 
\end{aligned} \right.
\end{equation}      
Problem $\mathcal{P}_5$ is still non-convex and the variable is coupled in both \ac{sct} and power constraints. To further simplify $\mathcal{P}_5$, we introduce an auxiliary variable $K$ representing the number of the independent sources that contribute to the \ac{rd} function. Namely, we have $g_i(\mathbf{p}) \ge \xi, i=1,2,\cdots, K$. Thus, $\mathcal{P}_5$ can be equivalently transformed into the following problem
\begin{equation} \label{pk40} 
\begin{aligned}
	\mathop { \text{max} }\limits_{\mathbf{p}, K} &\kern 15pt  \sum_{i=1}^{K} g_i(\mathbf{p})-K\xi   \\
	\text{s.t} \kern 5pt &  M_s \sum_{i=1}^{K} \log \frac{g_i(\mathbf{p})}{\xi} \le I_c(\mathbf{p}), \kern 2pt 0<K\le N, \\
	 &  \kern 15pt \mathbf{1}_N^T\mathbf{p} \le P_T, \kern 2pt \mathbf{p}\ge 0. 
\end{aligned}
\end{equation}
Note that $K$ can be immediately obtained through the reverse water-filling algorithm once the power vector $\mathbf{p}$ is determined.

\begin{algorithm}[!t]
	\caption{Modified gradient projection algorithm}
	\label{alg3}
	\textbf{Initialize:} The initial point $\mathbf{p}^{(0)} \in \mathcal{F}$, $\beta>0$, $\epsilon >0$. \\	
	\While{$|\tilde{h}(\mathbf{p}^{(l+1)})-\tilde{h}(\mathbf{p}^{(l)})|> \epsilon$}{
		Calculate $K^{(l)},\xi^{(l)}$ through reverse water-filling\; 
		Obtain the gradient $\nabla \tilde{h}(\mathbf{p}^{(l)})$ by \eqref{gradient}\;
		$\tilde{\mathbf{p}}^{(l)}=\text{Proj}(\mathbf{p}^{(l)}+\beta \nabla h(\mathbf{p}^{(l)}))$\;
		Find stepsize $\alpha^{(l)}$ based on the Armijo condition\;
		$\mathbf{p}^{(l+1)}=\mathbf{p}^{(l)}+\alpha^{(l)}(\tilde{\mathbf{p}}^{(l)}-\mathbf{p}^{(l)})$\;
	}
\end{algorithm}

\begin{figure}[!t]
	\centering
	\includegraphics[width= 3 in]{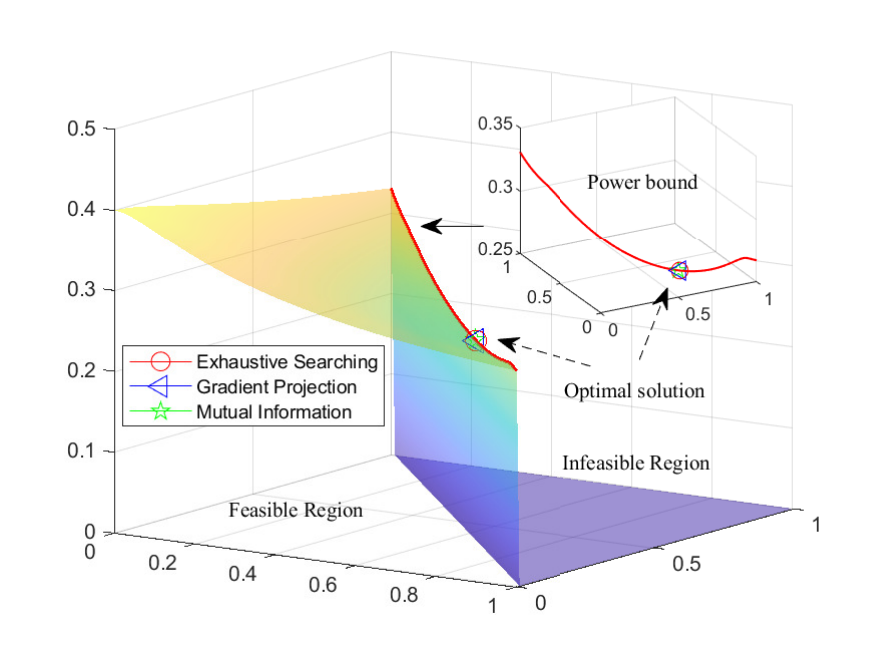}
	\caption{The visualization of the \ac{dw} waveform design for the case of $N=M_s=M_c=2$, $\text{SNR}_s=15$ dB, $\text{SNR}_c=0$ dB.}
	\label{2dexample}
\end{figure}

\noindent
\underline{\textbf{\textit{Proposition 3:}}} The SCT and power constraints are both tight for optimal solution in the DW design problem.  

\noindent   
\textbf{\textit{Proof:}} Please see Appendix \ref{apC}. $\hfill\blacksquare$

\begin{figure*}[!t]
	\centering
	\subfigure[ The case for SW design.]{
		\centering
		\includegraphics[width=0.65\linewidth]{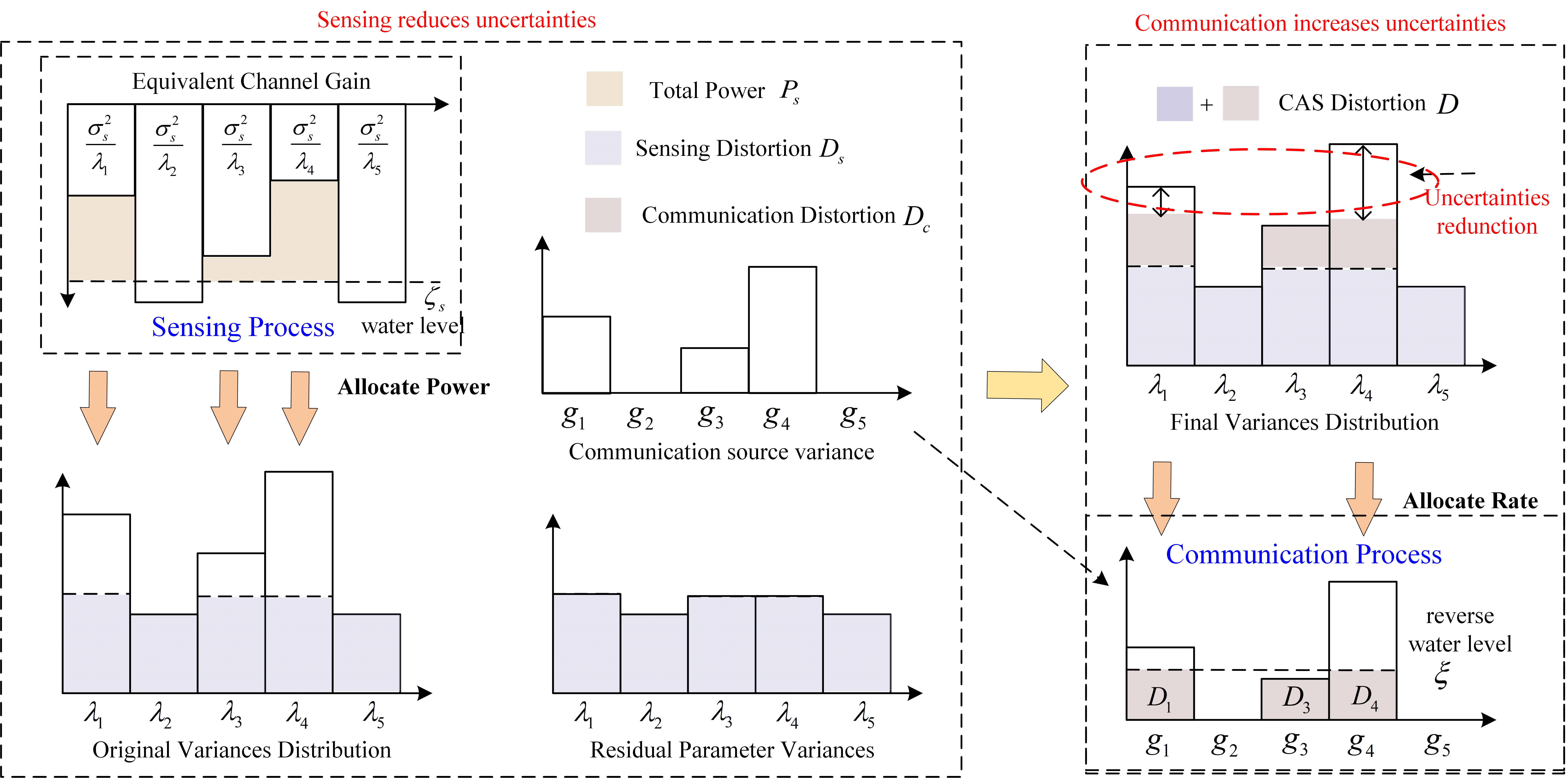} \label{New_Waterfill}}	
	\hspace{5mm}	
	\subfigure[The case for DW design.]{
		\centering
		\includegraphics[width=0.25\linewidth,height= 0.31\linewidth]{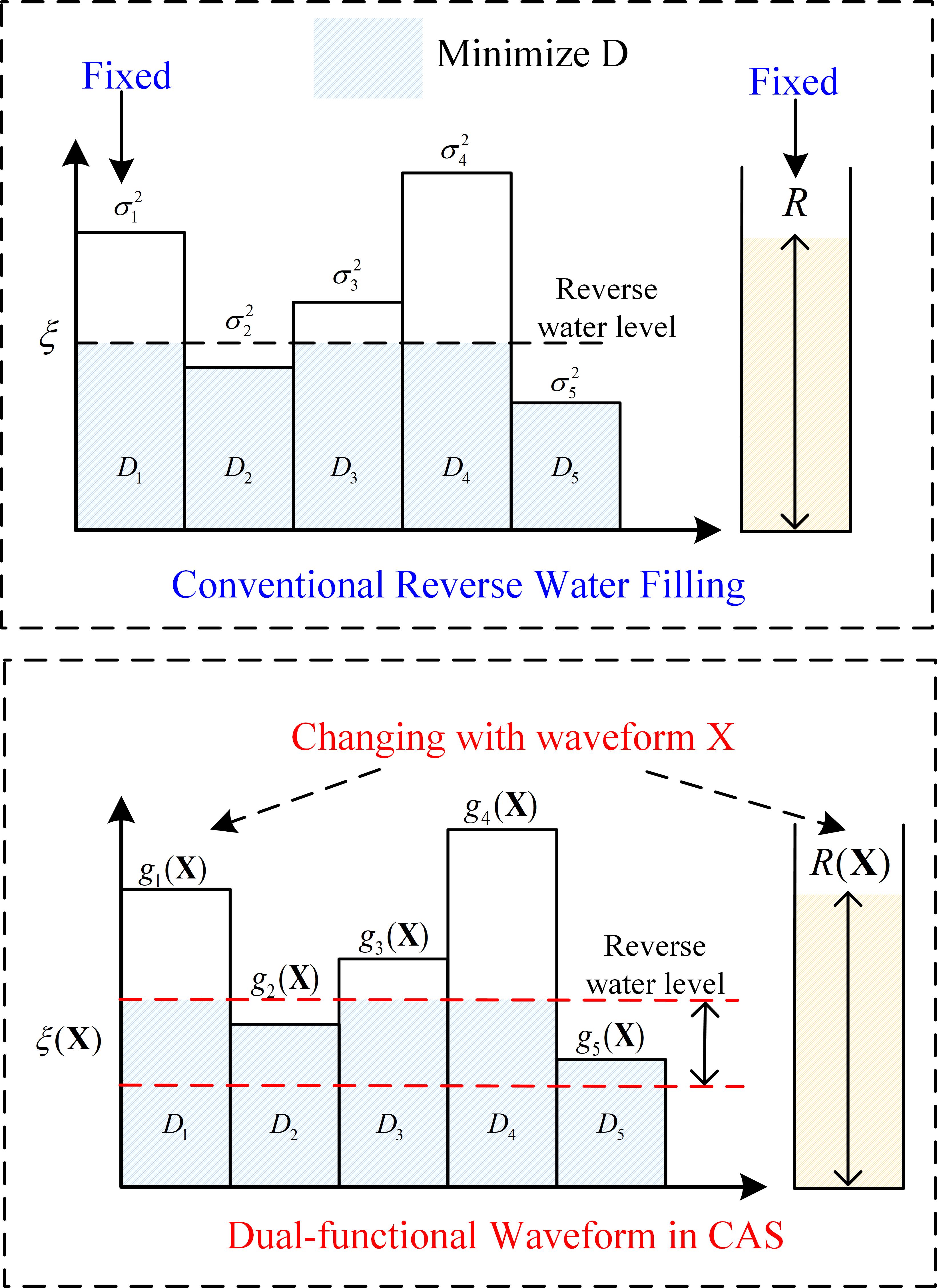}\label{New_Waterfill_Dual_function}}
	
	\caption{The variation in parameter uncertainties: (a) Uncertainties decrease in the SP while increasing in the CP, necessitating a well-designed waveform to reduce CAS distortion; (b) The DW scheme introduces a novel reverse water-filling problem.}
	\label{DiscussSWDW}
\end{figure*}

According to Proposition 3, $\xi$ can be expressed through the equality constraint, which is          
\begin{equation}
\xi = e^{\tilde{f}(\mathbf{p})}, \kern 2pt  \tilde{f}(\mathbf{p}) \triangleq \left(\sum_{i=1}^{K} \log g_i(\mathbf{p})-\frac{I_c(\mathbf{p})}{M_s}\right)/K. 
\end{equation}  
For a fixed $K$, problem \eqref{pk40} can be transformed into  
\begin{equation} \label{dd42} 
\mathop { \text{max} }\limits_{\mathbf{p} \in \mathcal{F}} \kern 5pt \tilde{h}(\mathbf{p}):=\sum_{i=1}^{K} g_i(\mathbf{p})-Ke^{{\tilde{f}(\mathbf{p})}},
\end{equation}
where the feasible region is denoted by $\mathcal{F} = \big\{\mathbf{p}|\mathbf{1}_N^T\mathbf{p} \le P_T, \mathbf{p}\ge 0 \big\}$. Since $\mathcal{F}$ is convex and the objective function is differentiable, we propose a gradient projection method to solve problem \eqref{dd42}. Specifically, we have 
\begin{equation}
\frac{\partial \tilde{h}(\mathbf{p})}{\partial p_i} = \frac{\lambda_{s_i}^2\sigma_s^2}{(\sigma^2_s+\lambda_{s_i}p_i)^2}-Ke^{\tilde{f}(\mathbf{p})}\frac{\partial \tilde{f}(\mathbf{p})}{\partial p_i},
\end{equation}    
where
\begin{equation}
\frac{\partial \tilde{f}(\mathbf{p})}{\partial p_i} = \frac{1}{K}\Big(\frac{\sigma^2_s}{\sigma^2_sp_i+\lambda_{s_i}p^2_i}-\frac{1}{M_s}\frac{\lambda_{c_i}}{\sigma_c^2+\lambda_{c_i}p_i}\Big). \nonumber
\end{equation}
Thus, the gradient of the objective $ \nabla \tilde{h}(\mathbf{p})$ can be obtained by
\begin{equation}\label{gradient}
\nabla \tilde{h}(\mathbf{p}) = \left \{
\begin{aligned}
\frac{\partial \tilde{h}(\mathbf{p})}{\partial p_i}, \kern 5pt &i = 1,2,\cdots, K, \\
0, \kern 15pt &i=K+1,\cdots,N.
\end{aligned}
\right.
\end{equation}  
Denote $\mathbf{p}^{(l)}$ as the feasible solution at the $l$-th iteration. While going a step forward along with the gradient direction, i.e., $\bar{\mathbf{p}}^{(l)}=\mathbf{p}^{(l)}+ \beta \nabla \tilde{h}(\mathbf{p}^{(l)})$ may improve the objective value, $\bar{\mathbf{p}}^{(l)}$ may not fall in the feasible set. The projection step onto the feasible set is essentially solving the following convex optimization problem 
\begin{equation}
\tilde{\mathbf{p}}^{(l)} =\text{Proj}(\bar{\mathbf{p}}^{(l)}) = \mathop { \arg \min }\limits_{\mathbf{x} \in \mathcal{F}} \kern 2pt \|\mathbf{x}-\bar{\mathbf{p}}^{(l)}\|.
\end{equation}
Thus, the feasible solution at the $(l+1)$-th iteration can be obtained by
\begin{equation}
\mathbf{p}^{(l+1)}=\mathbf{p}^{(l)}+\alpha^{(l)}(\tilde{\mathbf{p}}^{(l)}-\mathbf{p}^{(l)}).
\end{equation}
where $\alpha^{(l)}$ is the step size. By doing so, the optimal solution may be obtained after several iterations. The detailed algorithm procedure is summarized in Algorithm \ref{alg3}. 

To visualize the effectiveness of the two proposed \ac{dw} design algorithms, we take a low-dimensional case $N=M_s=M_c=2$ as an example. In such case, we can leverage exhaustive searching method to find the global optimal solution. To be specific, we collect the objective values over the 2-D grids $p_1 \times p_2$, where $p_1, p_2 \in [0, P_T]$. Fig. \ref{2dexample} verify the Proposition 1 and 3 that the optimal solution is achieved at the power bound, and the proposed two algorithms, namely, MI maximization and gradient projection both attain the optimal solution.

\begin{figure*}[!t]
	\subfigure[ The \ac{iid} scenario with $\lambda_s=1$, $\text{SNR}_s=20$ dB.]{
		\centering
		\includegraphics[width=0.32\linewidth]{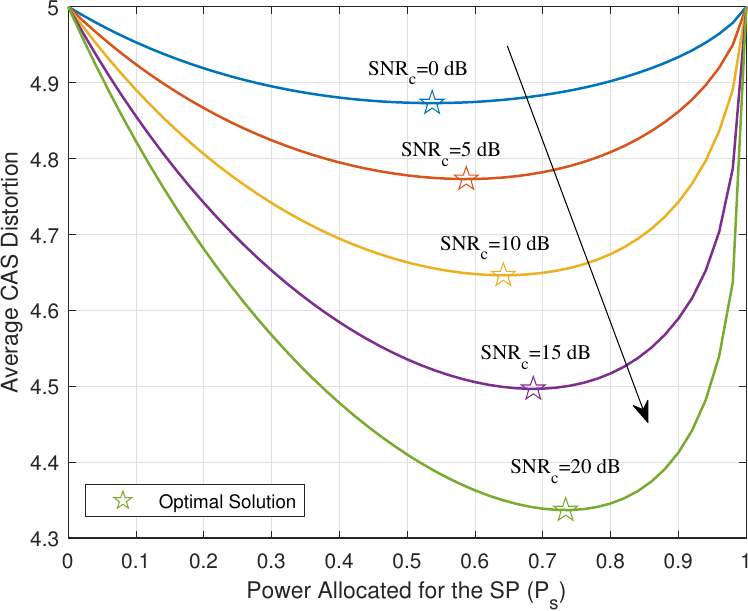} \label{f5a}}
	\subfigure[The general variance matrix, $\text{SNR}_c=20$ dB.]{
		\centering
		\includegraphics[width=0.32\linewidth]{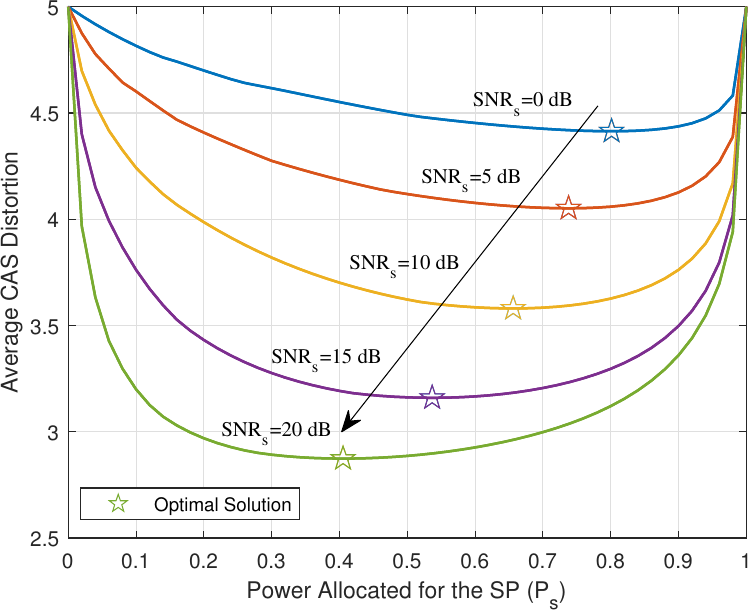}\label{f5b}}
	\subfigure[The power allocations corresponding to (b).]{
		\centering
		\includegraphics[width=0.32\linewidth]{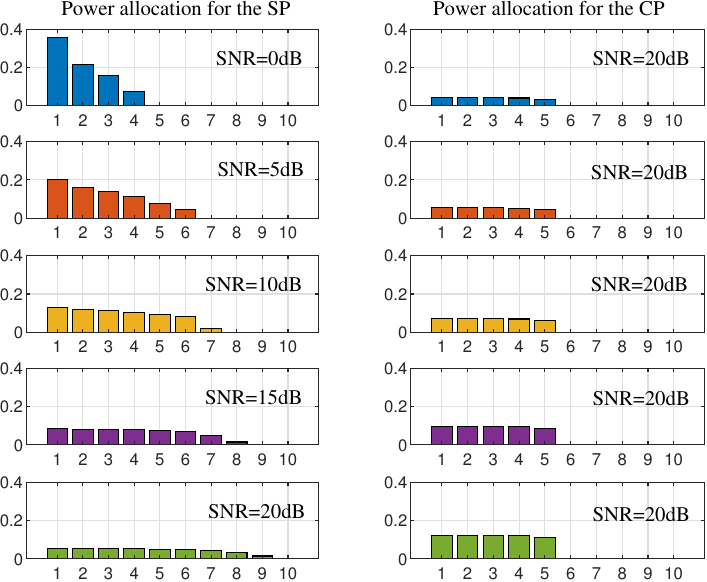} \label{f5c}}
	\caption{The illustration of the \ac{sw} design strategy: (a) Sensing \ac{qos} versus the power $P_s$ for the \ac{iid} sensing subchannels; (b) Sensing \ac{qos} versus the power $P_s$ for the general sensing variance matrix; (c) Detailed power distributions among the subchannels.}
	\label{scidentical}
\end{figure*}

\section{Discussion} \label{discussion}

In this section, we will provide insights into the \ac{cas} waveform design by ``tracking'' the parameter's uncertainties (i.e., variances) during the whole procedure. The original or prior variances of the equivalent independent parameters are denoted by $\bm{\lambda}=[\lambda_1, \lambda_2,\cdots, \lambda_N]$, which are the eigenvalues of the covariance matrix $\bm{\Sigma}_s$.    

\textit{The \ac{sw} scheme:} As shown in Fig. \ref{New_Waterfill}, in the SP, waver-filling algorithm is implemented to achieve the \ac{mmse}. Here, the parameter variances are divided into two categories: 1) Small variances which remain constant (no power allocation) throughout the \ac{cas} procedure, and 2) relatively large variances where the \ac{mmse} is attained through power allocation, resulting in a variance of $\sigma_s^2/\zeta_s$. The water-filling factor $\zeta_s$ is determined by the total power $P_s$ allocated for the SP. Consequently, we say that the SP can reduce the uncertainties of the parameters.

In the CP, the variances of the sources (estimates) are calculated by subtracting the \ac{mmse} from the original variances. The reverse water-filling algorithm is employed to minimize the communication distortion, with the reverse water-filling factor being chosen according to the achievable rate $I_c(P_c)$. Indeed, the CP may increase the uncertainties of the parameter acquired by sensing. In other words, the optimal \ac{mmse} estimation is attainable if the estimates can be transmitted losslessly, which implies a infinite channel capacity. We may observe that the final reduction of parameter uncertainties through the \ac{cas} equals to the reduction of the source variances during the CP, i.e., $g_i-\xi$. 

\textit{The \ac{dw} scheme:} As depicted in Fig. \ref{New_Waterfill_Dual_function}, the \ac{dw} design introduces a novel yet more intricate reverse water-filling problem. In the conventional scheme, the purpose of reverse water-filling is to minimize distortion with fixed variances and a given information rate. However, in the \ac{dw} scheme, the SP and CP are coupled \ac{wrt} the same waveform $\mathbf{X}$. This leads to the variances of the sources (estimates) $\lambda_i(\mathbf{R}_{\upeta})$ related to $\mathbf{X}$ and the \ac{mi} $I_c(\mathbf{X})$ (or the equivalent reverse water-filling factor $\xi(\mathbf{X})$), evolve simultaneously. This inherent interdependency is the essential challenge in finding an optimal solution for the \ac{dw} scheme.
       
\textbf{Remark 4:} In this paper, we exclusively focus on the comprehensible power allocation problem within the CAS systems. It is crucial to emphasize that system resources, such as bandwidth and dwell time, may also influence the SW and DW strategies, leading to performance tradeoffs distinct from those associated with power allocation \cite{9729751}. Furthermore, the DW is assumed to follow a Gaussian distribution, consistent with current signaling strategy in communication systems. However, this distribution may not be optimal for the CAS system since the deterministic waveforms are generally preferred for sensing. Investigating the optimal distribution and developing more efficient algorithms for the \ac{dw} design are also interesting topics, that are left for our future research.

\section{Simulation Results} \label{simulation}
In this section, we evaluate the effectiveness of the proposed waveform design strategies through numerical simulations. Unless otherwise specified, the system parameters are set as follows. The BS is equipped with $N=10$ transmitting antennas, while the number of receiving antennas for the SP and CP are set as $M_s=M_c=5$. The total number of transmit symbols is $T=100$. The communication channel is modeled by Rayleigh fading, where each entry of $\textbf{H}_c$ obeys the standard complex Gaussian distribution. Additionally, the sensing channel variance matrix is generated by \cite{8579200}
\begin{equation}
\mathbb{E}[\mathbf{h}\mathbf{h}^H]=\mathbf{I}_{M_s} \otimes \bm{\Sigma}_s=\mathbf{I}_{M_s} \otimes \sum_k \delta_k^2 \mathbf{a}(\theta_k)\mathbf{a}^H(\theta_k),   
\end{equation}
where $\mathbf{a}(\theta_k)=\frac{1}{\sqrt{N}}[1, e^{j \pi \sin (\theta_k)} \cdots, e^{j \pi N \sin (\theta_k)}]$ and $\delta_k$ represent the transmitting array steering vector and the expectation of the path-loss for the $k$-th path, respectively. Without loss of generality, we set $\delta_k =1$ and randomly choose $K=10$ angles over the interval $[-90^\circ,90^\circ]$. Moreover, we use the normalized transmit power $P_T = T$ and define the \ac{snr} by $\text{SNR}=10\log(P_T/\sigma^2)$ dB.   

\subsection{Evaluation of the \ac{sw} Design}
In the scenario of the \ac{sw} design, the power distribution among the subchannels and objective value are determined with a given power $P_s$ allocated for the SP, or equivalent power $P_T-P_s$ for the CP. Therefore, we illustrate the tradeoff in performance between \ac{snc} by presenting the curves of the CAS distortion versus power allocated for the SP, where the average CAS distortion is defined by $D/N$.

Let us begin by examining the specific case of \ac{iid} sensing subchannels with $\lambda_s = 1$. The SNR for the sensing channel is set at a constant value of $\text{SNR}_s=20$ dB. Meanwhile, we vary SNR for the communication channel $\text{SNR}_c$ from 0 dB to 20 dB in increments of 5 dB. At first glance, Fig. \ref{f5a} confirms the statement in Proposition 2 that the objective function is convex \ac{wrt} $P_s$. Therefore, the proposed one-dimensional search algorithm can effectively find the optimal solution. On one hand, we observe a discernible \emph{power allocation tradeoff}. This tradeoff manifests when allocating excessive power to either side, resulting in significant performance degradation, until the achieved distortion is equal to the sum of prior variances. On the other hand, more power tends to be allocated to the SP as the communication \ac{snr} level increases. This observation suggests that the BS tends to acquire more accurate sensing data when the communication quality is sufficiently high.                

In Fig. \ref{f5b}, we present the performance tradeoff analysis for the general case, where the sensing variance matrix is generated by random angles. To eliminate redundancy, within this simulation, we maintain a fixed communication \ac{snr} at 20 dB, while varying the sensing \ac{snr} from 0 dB to 20 dB. Similar to the observations in Fig. \ref{f5a}, it is evident that the CAD distortion decreases as the sensing SNR increases. It is noteworthy that when the sensing SNR reaches a sufficiently high level, the BS expends a greater portion of its resources to enhance the communication capabilities. Fig. \ref{f5c} provides the detailed power distribution among the \ac{snc} subchannels, where the channel qualities are arranged in descending order. We can observe that, in situations characterized by low SNR, a higher proportion of power is allocated to the subchannels with superior quality, while power allocation tends to become more uniform as the SNR reaches sufficiently high levels.  

\begin{figure}[!t]
	\centering
	\includegraphics[width=3in]{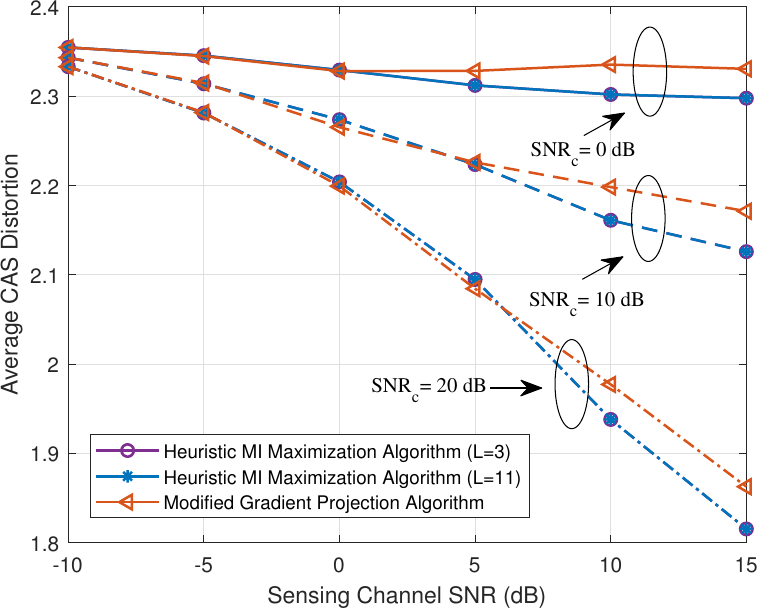}
	\caption{The CAS distortion versus various \ac{snc} channel SNRs in the scenario of the independent sensing subchannels.}
	\label{f6a}
\end{figure}


\subsection{Evaluation of the \ac{dw} Design}

In this subsection, we evaluate the CAS performance for the \ac{dw} signaling strategy. For comparison purpose, we present the performance of heuristic MI maximization (HMI) algorithm with two sets of weighting factors: one with $L=3$ grids encompassing three special solutions, i.e.,  the communication-optimal ($\alpha = 0$), the sensing-optimal ($\alpha = 1$), and the sum of \ac{snc} MI ($\alpha = 0.5$); the other with $L=11$ grids to explore a wider range of weight combinations.

In Fig. \ref{f6a}, we provide the CAS distortion curves under varying \ac{snc} SNRs for independent sensing subchannels. In this scenario, the eigenvalues of the covariance matrix $\bm{\Sigma}_s$ are extracted to form a new diagonal covariance matrix. Note that the modified gradient projection (MGP) algorithm is specifically designed for this scenario. We can observe three key phenomena as follows. 1) There is noticeable trend of improved sensing \ac{qos} with increasing \ac{snc} channel SNRs. 2) The HMI algorithm outperforms the MGP algorithm, due to the sensitivity of the MGP algorithm to the convexity of the objective function and the choice of the initial point. However, as observed in the 2D example, the objective function is not necessarily convex, resulting in a performance loss. 3) Interestingly, the HMI algorithm exhibits the same performance for both two sets of weight factors. This consistency can be attributed to the optimal weight factor, as detailed in Table \ref{table1}. 
  
Surprisingly, the optimal weight factors for both sets consistently result in values of either 0 or 1 in this experiment. It implies that the optimal solution is achieved at either the sensing-optimal or the communication-optimal point. This phenomenon can be attributed to the rearrangement of the eigenvalues of both the \ac{snc} subchannels, e.g., in descending order, to align the qualities of the \ac{snc} subchannels consistently. This operation can be accomplished by adjusting the eigenvectors thanks to the absence of the \emph{subspace trade-off} in this case. Accordingly, optimizing communication performance simultaneously improves the sensing performance, and vice versa. Furthermore, either the SP or CP with relatively poorer channel quality takes precedence for optimization, ensuring compliance with the \ac{sct} constraint.

\begin{table}[!htbp] 
	\centering
	\caption{The optimal weighting factor $\alpha$.}      
	\label{table1}
	\begin{tabular}{ccccccc}
		\toprule
		\diagbox{$\text{SNR}_c$}{$\text{SNR}_s$}& -10 dB & -5 dB & 0 dB & 5 dB & 10 dB & 15 dB  \\
		\midrule
		0 dB  & 1 &1 & 0 & 0 & 0 & 0  \\ 
		\midrule	
		10 dB & 1 &1 & 1 & 0 & 0 & 0 \\ 
		\midrule
		20 dB & 1 &1 & 1 & 1 & 0 & 0  \\ 
		\bottomrule
	\end{tabular}
\end{table} 

\begin{figure}[!t]
	\centering
	\includegraphics[width=3in]{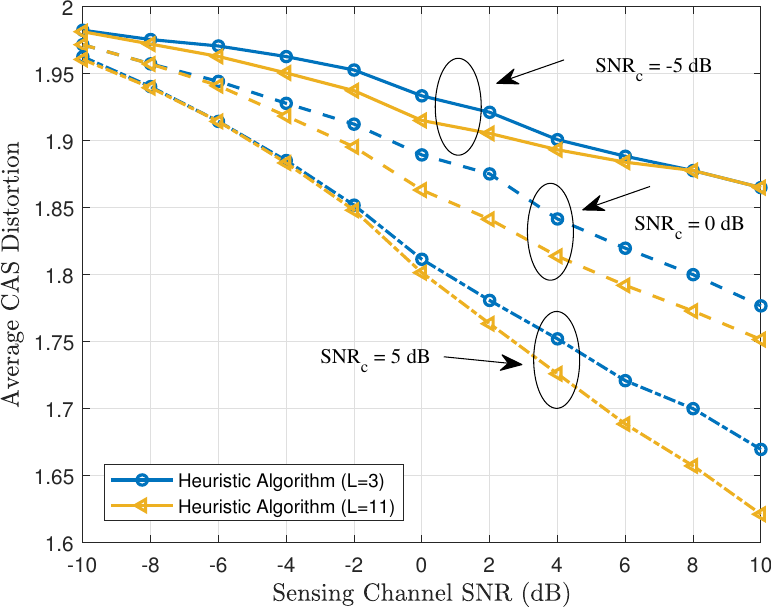}
	\caption{The CAS distortion versus various \ac{snc} channel SNRs for the general sensing covariance matrix.}
	\label{DW_HighD}
\end{figure}

In Fig. \ref{DW_HighD}, we evaluate the CAS performance for a general covariance matrix $\bm{\Sigma}_s$, with sensing SNR ranging from -10 dB to 10 dB with spacing 2 dB. It is important to note that the MGP algorithm is not applicable in this context. Additionally, the eigenvalue rearrangement operation for the sensing covariance matrix is ineffective due to the presence of \ac{snc} subspace tradeoff. Consequently, for the HMI algorithm, refining the grid resolution of the weighting factor $\alpha$ becomes crucial for enhancing CAS performance, as evidenced by the performance gap between $L=3$ and $L=11$ shown in Fig. \ref{DW_HighD}. Notably, the performance of these two algorithms appears to converge under certain extreme conditions. For instance, the CAS optimal solution may coincide with the communication-optimal solution when the sensing channel quality is relatively favorable, such as at $\text{SNR}_s$ = 10 dB, $\text{SNR}_c$ = -5 dB. Conversely, it may align with the sensing-optimal solution for the relatively favorable communication channel quality, e.g., at $\text{SNR}_s$ = -10dB, $\text{SNR}_c$ = 5dB.
            
\subsection{Comparison of the \ac{sw} and \ac{dw} Designs}

Finally, we conduct a comparative analysis of the performance between the \ac{sw} and \ac{dw} designs within the context of \ac{cas} systems, while keeping the system conditions unchanged. To encompass a wide range of sensing channel qualities, we vary the sensing SNR from -20 dB to 50 dB. In addition to the \ac{sw} and \ac{dw}, we provide a benchmark scheme of the \ac{dw} with independent sensing subchannel (DWIS), where the variances of the sensing subchannel are equal to the eigenvalues of $\mathbf{R}_s$. By providing the simulations of DWIS, we aim to demonstrate the impact of the \emph{subspace tradeoff}. 

\begin{figure}[!t]
	\centering
	\includegraphics[width=3in]{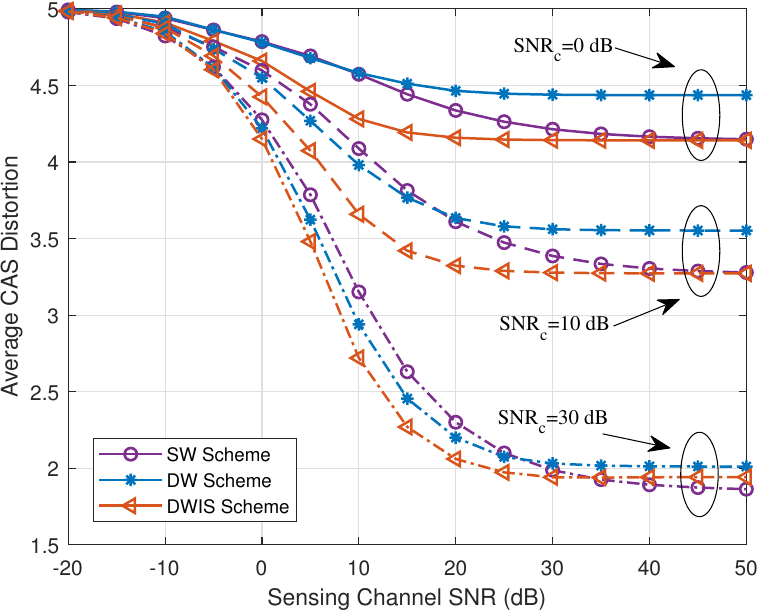}
	\caption{The comparison of the \ac{sw} and \ac{dw} schemes.}
	\label{Compare_SD}
\end{figure}

As depicted in Fig. \ref{Compare_SD}, it is evident that the DWIS scheme outperforms its counterparts. This superiority is attributed to the fact that the eigenspace of the waveform aligns perfectly with the optimal \ac{snc} subspaces, while simultaneously attaining a power multiplexing gain. Interestingly, the intersection points are observed between the \ac{sw} and \ac{dw} schemes in the moderate SNRs. In the regime of low SNRs, the power allocation between the SP and CP dominates the CAS performance. Therefore, the \ac{dw} outperforms the \ac{sw}, benefiting from the resource multiplexing gains. Conversely, the performance of the \ac{sw} surpasses the \ac{dw} strategies when the \ac{snc} channels exhibit sufficiently high quality. As resource multiplexing gains approach saturation levels, the optimal waveform structures take precedence in influencing the CAS performance. In the \ac{sw} scheme, both \ac{snc} sides can optimized simultaneously. This stands in contrast to the \ac{dw} scheme due to the existence of \emph{water-filling tradeoff}. Furthermore, the reordering of \ac{snc} subchannels qualities into a consistent order becomes unfeasible due to the presence of the \emph{subspace tradeoff}, resulting in a further performance loss.

\section{Conclusion}\label{Conclusion}
In this article, we developed a communication-assisted sensing (CAS) framework in 6G perceptive networks. Our framework is grounded in the principles of the rate-distortion theory and the source-channel separation theory in lossy data transmission. We conceived efficient waveform design algorithms tailored for two typical signaling strategies, namely, the separated sensing and communication (S\&C) waveform and dual-function waveform, within the context of target response matrix estimation. Furthermore, we have conducted in-depth discussions of tradeoffs in power allocation for the separated \ac{snc} waveform, and the tradeoffs in subspace and water-filling for the dual-functional waveform. Our numerical simulations confirmed the effectiveness of the proposed algorithms. Meanwhile, the results highlight that the dual-function scheme outperforms the separated \ac{snc} scheme in the scenarios characterized by relatively poor channel conditions, as it capitalizes on the resource multiplexing gains. Conversely, the separated \ac{snc} scheme emerges as the preferable choice when the channel qualities reach a sufficiently high level.     
    
\begin{appendices}  
	
\section{}\label{apA} 

\subsubsection{Tightness of the \ac{sct} Constraint} Assume that $(\mathbf{p}_s^\star,\mathbf{p}_c^\star)$ is the optimal solution such that $R(D_c)<I_c$. The variance of the source distribution $g_i(p_{s_i})$ and the achievable rate $I_c$ are determined by $(\mathbf{p}_s^\star,\mathbf{p}_c^\star)$. Then, we can always improve the coding rate by appropriately choosing the reverse water-filling factor $\xi$ until reaching the communication achievable rate, namely, $R(\tilde{D}_c)=I_c$. By recalling the non-increasing property of RD function, we have $ \tilde{D}_c \le D_c $ due to $R(\tilde{D}_c) = I_c > R(D_c)$. Note that adjusting the reverse water-filling factor $\xi$ is irrelevant to the variables $\mathbf{p}_s$ and $\mathbf{p}_c$, but it may further reduce the objective value. 

\subsubsection{Tightness of the Power Constraint} Assume that $(\mathbf{p}_{s}^\star,\mathbf{p}_c^\star)$ is the optimal solution such that $\mathbf{1}_N^T\big(\mathbf{p}_{s}^\star+\mathbf{p}_{c}^\star\big) < P_T$. Let us denote $\Delta P$ as the residual power such that $\mathbf{1}_N^T\big(\mathbf{p}_{s}^\star+\mathbf{p}_{c}^\star\big) + \Delta P = P_T$. Then, we can always allocate the residual power to the CP and obtain the new water-filling solution $\tilde{\mathbf{p}}_{c}^\star$ satisfying
\begin{equation}
\mathbf{1}_N^T\tilde{\mathbf{p}}_{c}^\star = \mathbf{1}_N^T\mathbf{p}_{c}^\star+\Delta P.
\end{equation}                     	
Then we have $I_c(\tilde{\mathbf{p}}_{c}^\star) \ge I_c(\mathbf{p}_{c}^\star)$ due to the fact that the MI is a monotonically non-decreasing function \ac{wrt} the SNR. Similarly, a smaller $\tilde{D}_c$ can be attained by adjusting the reverse water-filling factor $\xi$ such as $R(\tilde{D}_c)=I_c(\tilde{\mathbf{p}}_{c}^\star)$. Consequently, the values of objective satisfies 
\begin{equation}
D_s(\mathbf{p}_{s}^\star)+D_c(\mathbf{p}_{c}^\star) \ge D_s(\mathbf{p}_{s}^\star)+\tilde{D}_c(\tilde{\mathbf{p}}_{c}^\star),
\end{equation}     	
which is contradicted to the assumption that $(\mathbf{p}_{s}^\star,\mathbf{p}_c^\star)$ is the optimal solution. This completes the proof. $\hfill\blacksquare$
	
\section{}\label{apB} 
Let us denote the objective function as
\begin{equation}
h(x) = \big(1-e^{-\tilde{I}(x)}\big)f(x)+\lambda_se^{-\tilde{I}(x)}, \kern 2pt x \in [0,P_T].
\end{equation} 	
where $\tilde{I}(x) = \frac{I_c(P_T-x)}{M_sN}$. The first-order derivative of $h(x)$ can be expressed by
\begin{equation}
h'(x)=f'(x)+[f(x)\tilde{I}'(x)-\lambda_s\tilde{I}'(x)-f'(x)]e^{-\tilde{I}(x)}.
\end{equation} 	
Then, the second-order derivative is obtained by
\begin{equation}
\begin{aligned}
h''(x)&=(1-e^{-\tilde{I}(x)})f''(x)\\
      &+e^{-\tilde{I}(x)}(f(x)-\lambda_s)(\tilde{I}''(x)-(\tilde{I}'(x))^2).
\end{aligned}
\end{equation} 	
In the first term, we have $(1-e^{-\tilde{I}(x)}) >=0$ due to $\tilde{I}(x) \ge 0$ in the domain of $x$. $f''(x) \ge 0$ since $f(x)$ (i.e., the \ac{mmse}) is a convex function of $x$. In the second term, it is evident that $f(x)-\lambda_s = -g(x) \le 0$ according to the definition \eqref{dd26}. We also have $(\tilde{I}''(x)-(\tilde{I}'(x))^2) \le 0$ since $\tilde{I}''(x) \le 0$ due to the fact that $\tilde{I}(x)$ (i.e., the achievable rate) is the concave function of SNR \cite{1412024}. Consequently, we have $h''(x) \ge 0$ which indicates that $h(x)$ is a convex function. This completes the proof. $\hfill\blacksquare$

\section{}\label{apC} 

The tightness of the SCT constraint is analogous to that of the SW design, as adjusting the reverse water-filling factor $\xi$ is independent of the power allocation vector $\mathbf{p}$. However, the tightness of the power constraint is less straightforward. This complexity arises because changes to $\mathbf{p}$ influence both sides of the SCT inequality. 

Assume that $\mathbf{p}^\star$ is the optimal solution satisfying $\mathbf{1}_N^T \mathbf{p}^\star < P_T$. We demonstrate that it is always possible to construct a new vector $\mathbf{\tilde{p}}^\star$ such that $\mathbf{1}_N^T \mathbf{\tilde{p}}^\star = P_T$, potentially reducing the CAS distortion while meeting the new SCT constraint. To proceed, note that the following facts: 

\begin{itemize}
\item $I_c(\mathbf{\tilde{p}}^\star) \ge I_c(\mathbf{p}^\star)$, as MI is a monotonically non-decreasing function \ac{wrt} the SNR;

\item $f_i(\tilde{p}_i) \le f_i(p_i)$, thereby $D_s(\mathbf{\tilde{p}}^\star) \le D_s(\mathbf{p}^\star)$, as MMSE is a monotonically non-increasing function \ac{wrt} the SNR; 

\item $g_i(\tilde{p}_i) \ge g_i(p_i)$ due to the fact that $ g_i(p_i) = \lambda_{s_i} - f_i(p_i) $. 

\end{itemize} 

Next, consider the reverse water-filling procedure \eqref{rdcc} with the factor $\xi$ and power allocation vector $\mathbf{p}^\star$. For a positive coding rate $I_c(\mathbf{p}^\star)$, there must be some source variances satisfying $ g_i(p_i) > \xi $; otherwise, the coding rate would be zero according to the RD function. Let us define the index set $\mathcal{I}_1 = \{ i | g_i(p_i) > \xi \}$. Thus, we have $D_{c_i}(p_i) = \xi$ for $i \in \mathcal{I}_1$. By allocating the residual power $\Delta P = P_T - \mathbf{1}_N^T \mathbf{p}^\star$ to subchannels $i \in \mathcal{I}_1$, we get a reduction in MMSE but an equivalent increment in source variance, namely,  
\begin{equation}
f_i(\tilde{p}_i) = f_i(p_i) - \Delta D, \kern 5pt  g_i(\tilde{p}_i) = g_i(p_i) + \Delta D. 
\end{equation}
For discussion convenience, we consider the $i$-th subchannel without loss of generality. Note that
\begin{equation}\label{61}
\log \frac{g_i(p_i) + \Delta D}{\xi + \Delta D} \mathop \le \limits^{(a)} \log \frac{g_i(p_i)}{\xi} = I_c(\mathbf{p}^\star) \le I_c(\mathbf{\tilde{p}}^\star) 
\end{equation}
where $(a)$ holds because $f(x) = (a+x)/(b+x)$ is a decreasing function for $x>0$ when $a>b$. Formula \eqref{61} implies that there exists a new water-filling factor $\tilde{\xi}$ satisfying
\begin{equation}
\tilde{\xi} \le  \xi + \Delta D \le g_i(p_i) + \Delta D,
\end{equation}       
such that the new SCT constraint $\log g_i(\tilde{p}_i)/\tilde{\xi} = I_c(\mathbf{\tilde{p}}^\star) $ is met. In this scenario, the CAS distortion is given by
\begin{equation}
D(\tilde{p}_i) = D_s(p_i) - \Delta D + \tilde{\xi} \le D_s(p_i) + \xi = D(p_i).
\end{equation} 
In summary, by allocating the residual power to the subchannels $i \in \mathcal{I}_1$, the reduction in sensing distortion must be greater than or equal to the increase in communication distortion, leading to a potential reduction in overall CAS distortion. This completes the proof.  $\hfill\blacksquare$
  		
\end{appendices}

\bibliographystyle{IEEEtran}
\bibliography{IEEEabrv,CAS}

\end{document}